\documentclass[11pt,twoside,a4paper]{article}
\usepackage{color}
\usepackage{comment}
\usepackage{amsmath}
\usepackage{cleveref}
\usepackage{graphicx}
\graphicspath{{figures/}}
\usepackage[font=small,labelfont=bf]{caption}
\usepackage{authblk}
\usepackage{subcaption}
\usepackage{siunitx}
\usepackage{url}
\DeclareSIUnit\Molar{M}

\begin{document}
\title{Machine learning a model for RNA structure prediction}
\author[1]{Nicola Calonaci}
\author[2,3]{Alisha Jones}
\author[1]{Francesca Cuturello}
\author[2,3]{Michael Sattler}
\author[1]{Giovanni Bussi}
\affil[1]{International School for Advanced Studies, via Bonomea 265, 34136 Trieste, Italy}
\affil[2]{Institute of Structural Biology, Helmoltz Zentrum M\"{u}nchen, Neuherberg 85764, Germany}
\affil[3]{Center for Integrated Protein Science M\"{u}nchen and Bavarian NMR Center at Department of Chemistry, Technical University of Munich, Garching 85757, Germany}
\date{}
\maketitle

\begin{abstract}
RNA function crucially depends on its structure. Thermodynamic models currently used for secondary structure prediction
rely on computing the partition function of folding ensembles, and can thus estimate minimum free-energy
structures and ensemble populations.  
These models sometimes fail in identifying native structures
unless complemented by auxiliary experimental data.
Here, we build a set of models that combine thermodynamic parameters,
chemical probing data (DMS, SHAPE), and co-evolutionary data
(Direct Coupling Analysis, DCA) through a network that outputs perturbations to the ensemble free energy.
Perturbations are trained to increase the ensemble populations of a representative set of known native RNA structures.
In the chemical probing nodes of the network, a convolutional window combines neighboring reactivities, 
enlightening their structural information content and the contribution of local conformational ensembles.
Regularization is used to limit overfitting and improve transferability. The most transferable model is selected through a cross-validation strategy
that estimates the performance of models on systems on which they are not trained.
With the selected model we obtain increased ensemble populations for native structures
and more accurate predictions
in an independent validation set.
The flexibility of the approach allows the model to be easily retrained and adapted to incorporate arbitrary experimental information.
\end{abstract}

\sloppy

\section{Introduction}

Ribonucleic acids (RNA) transcripts, and in particular non-coding RNAs, play a fundamental role in cellular metabolism
being involved in protein synthesis \cite{cech2000ribosome}, catalysis \cite{doud-cech02nat}, and regulation of gene expression \cite{morris2014riseregRNA}.
RNAs often adopt dynamic interconverting conformations, to regulate their functional activity.
Their function is however largely dependent on a specific active conformation \cite{wan2011understanding}, making RNA structure determination
fundamental to identify the role of transcripts and the relationships between mutations and diseases \cite{cooper2009rna}.
The nearest-neighbor models based on thermodynamic parameters \cite{tinoco1973improved,andronescu2007nndb} allow
the stability of a given RNA secondary structure to be predicted with high reliability,
and dynamic programming algorithms \cite{nussinov1978algorithms,lorenz2011viennarna} can be used to quickly identify
the most stable structure or the entire partition function for a given RNA sequence.
However, the coexistence of a large number of structures in a narrow energetic range \cite{wuchty1999complete}
often makes the interpretation of the results difficult.
Whereas there are important cases where multiple structures are indeed expected to coexist \emph{in vivo} and might be necessary for function \cite{dethoff2012visualizing,serganov2013decade},
the correct identification of the dominant structure(s) is crucial to elucidate RNA function and mechanism of action.
In order to compensate for the inaccuracy of thermodynamic models, it is becoming common to complement them with chemical probing data
\cite{weeks2010advances}
providing nucleotide-resolution information that can be used to infer pairing propensities
(\textit{e.g.}, reactive nucleotides are usually unpaired).
Particularly interesting is selective 2$^{\prime}$ hydroxyl acylation analyzed via primer extension (SHAPE)
\cite{merino2005rna,deigan2009accurate}, as it can also probe RNA structure \emph{in vivo} \cite{spitale2015structural}.
In a separate direction, novel methodologies based on direct coupling analysis (DCA) have been developed to optimally exploit co-evolutionary information
in protein structure prediction \cite{morcos2011direct} and found their way in the RNA world as well \cite{de2015direct,weinreb20163d}.
Whereas the use of chemical probing data and of multiple sequence alignments in RNA structure prediction is becoming more and more common, these two types 
of information have been rarely combined \cite{lavender2015shapemsa}.

In this paper, we propose a model to optimally integrate RNA thermodynamic models,
chemical probing experiments, and DCA co-evolutionary information
into a robust structure prediction protocol.
A machine learning procedure is then used to select the appropriate model and optimize the model parameters based on available experimental structures.
Regularization hyperparameters are used to tune the complexity of the model thus controlling overfitting and enhancing transferability.
The resulting model leads to secondary structure prediction
that surpasses available methods when used 
on a validation set not seen in the training phase.
The parameters can be straightforwardly re-trained on new available data.

\newpage

\section{MATERIALS AND METHODS}

\begin{figure*}
\centering
    \begin{subfigure}[t]{0.5\textwidth}
        \includegraphics[width=1.0\linewidth, keepaspectratio]{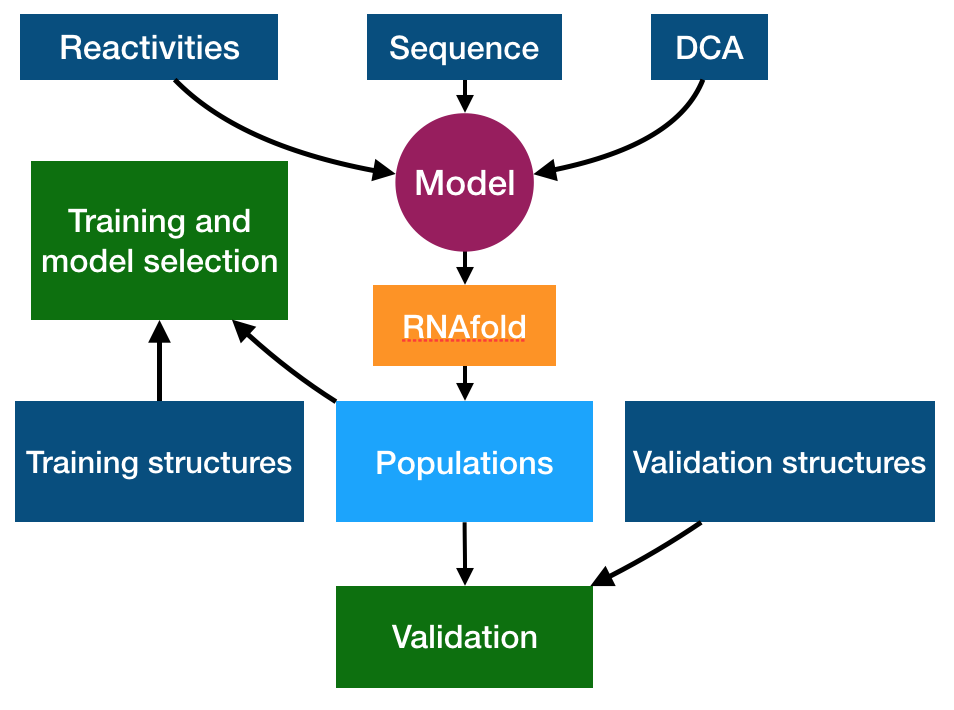}
        \caption{}
    \end{subfigure}\hfill
    \begin{subfigure}[t]{0.5\textwidth}
        \includegraphics[width=1.0\linewidth]{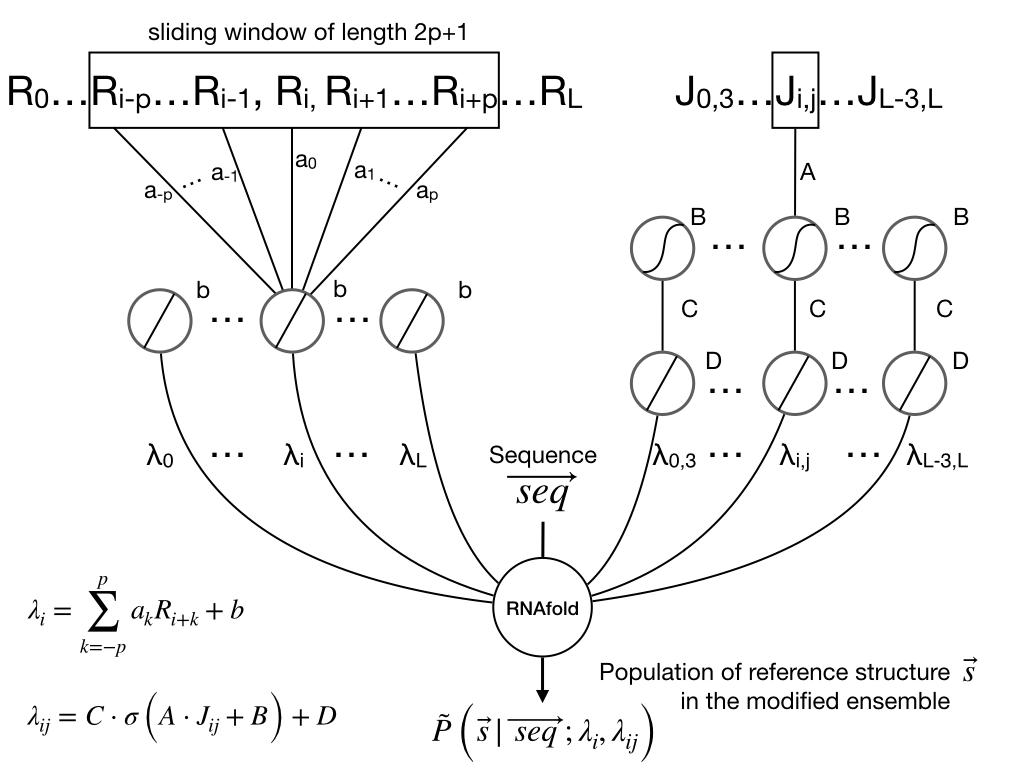}
        \caption{}
    \end{subfigure}
    \caption{Graphical scheme of the machine learning procedure. 
 (a) Models that integrate RNAfold, chemical probing experiments, and DCA scores into prediction of structure populations 
     are trained. One among all the proposed models is selected based on a transferability criterion and validated
     against data that is not seen during training. Available reference structures are used as target for training and validation.
 (b) Sequence, reactivity profile, and DCA data are included through additional terms in the RNAfold model free energy. 
 The network is split into two channels: a
 single-layered channel for reactivity input (left side) and a double-layered channel for DCA couplings (right side). Along the reactivity channel, a convolutional layer
 operates a linear combination on the sliding window including the reactivity $R_{i}$ of a nucleotide and the reactivities $\{R_{i+k}\}$ of its neighbors,
 with weights $\{a_k\}$ and bias $b$.
 The output consists in a pairing penalty $\lambda_i$ for the $i$-th nucleotide.
 In the DCA channel, the first layer transforms the input DCA coupling $J_{ij}$ via a non-linear (sigmoid) activation function, with weight $A$ and bias $B$.
 The transformed DCA input is then mapped to a pairing penalty $\lambda_{ij}$ for the specific $ij$ pair via a second layer, implementing a linear activation
 function with weight $C$ and bias $D$.
 Penalties for both individual nucleotides and for specific pairs are applied as perturbations to the RNAfold free-energy model.}
    \label{netscheme}
\end{figure*}
The architecture of the model is summarized in Fig.~\ref{netscheme}.

\subsection{Secondary structure annotation}

The secondary structures that we use for training and validation were obtained by
annotating crystallographic structures with \texttt{x3dna-dssr} \cite{lu2015dssr}.
Differently from previous work, we include all the computed cis-Watson-Crick contacts as reference base pairs, with exception of pseudoknots
that are forbidden in predictions made with RNAfold.
All of the reference structures are published in the PDB database and
have a resolution better than \SI{4}{\angstrom} so that they can be assumed to be of similar quality,
although crystal packing effects or other artefacts might in principle be different.
The list of PDB files used in this work is reported in Table \ref{dataset}.
\begin{table}
\centering
\begin{tabular}{l l l l l l l l}
 Molecule & PDB & $\textit{l}_{seq}$ & S1--S4\\ 
\hline
yeast Phe-tRNA & 1EHZ & 76 & \texttt{VTTT}\\                         
D5,6 Yeast ai5g G-II Intron & 1KXK & 70 & \texttt{TTTT}\\
Ribonuclease P RNA & 1NBS & 150 & \texttt{TTTT}\\
Adenine riboswitch & 1Y26 & 71 & \texttt{VTTV}\\
TPP riboswtich & 2GDI & 78 & \texttt{TTVT}\\
SAM-I riboswitch & 2GIS & 94 & \texttt{TTVT}\\
Lysine riboswitch & 3DIG & 174 & \texttt{TTVV}\\
O. I. G-II Intron & 3IGI & 388 & \texttt{TVTV}\\
c-di-GMP riboswitch & 3IRW & 90 & \texttt{TTTT}\\
M-box riboswitch & 3PDR & 161 & \texttt{VVTT}\\
THF riboswitch & 3SD3 & 89 & \texttt{VVTT}\\
Fluoride riboswitch & 3VRS & 52 & \texttt{VTTT}\\
SAM-I/IV riboswitch & 4L81 & 96 & \texttt{TVVV}\\
Lariat capping ribozyme & 4P8Z & 188 & \texttt{TTTT}\\
ydaO riboswitch & 4QLM & 108 & \texttt{TTTT}\\
ZMP riboswitch & 4XW7 & 64 & \texttt{VVVV}\\
50S ribosomal & 4YBB\_CB & 120 & \texttt{TVVV}\\
5-HTP RNA aptamer & 5KPY & 71 & \texttt{TTTT}\\
\hline
\end{tabular}
\caption{RNA molecules included in the dataset. For each molecule we indicate the PDB ID of the corresponding 
 annotated structure, the number of nucleotides ($\textit{l}_{seq}$), and, for each random dataset splitting that we used (S1 to S4), a mark to denote whether 
 the molecule data are used for training (T) or validation (V).
For PDB 4YBB, chain CB was used as a reference.}
\label{dataset}
\end{table}

\subsection{Thermodynamic model}

As a starting point we use the nearest neighbor thermodynamic model \cite{tinoco1973improved,andronescu2007nndb}
as implemented using dynamic programming \cite{nussinov1978algorithms} in the ViennaRNA package \cite{lorenz2011viennarna}.
Given a sequence $\vec{seq}$ the model estimates the free energy associated to any possible
secondary structure $\vec{s}$ by means of a sum over consecutive base pairs, with parameters based on the identity
of each involved nucleobase.
We denote this free energy as $F_0\left(\vec{s} | \vec{seq}\right)$.
We used here the thermodynamic parameters derived in \cite{andronescu2007nndb} as implemented in the ViennaRNA package, but the method could be
retrained starting with alternative parameters.
The probability $P_0\left(\vec{s} | \vec{seq}\right)$ of a structure $\vec{s}$ to be observed is thus
$
P_0\left(\vec{s} | \vec{seq}\right)=e^{-\frac{1}{RT}F_0\left(\vec{s} | \vec{seq}\right) }/Z_0\left(\vec{seq}\right)
$
where $Z_0$ is the partition function, $R$ is the gas constant and $T$ the temperature, here set to 300K.
Importantly, the implemented algorithm is capable of finding not only the most stable structure associated to a sequence ($\text{arg min}_{\vec{s}} F_0\left(\vec{s} | \vec{seq}\right)$)
but also the full partition function $Z_0$ and the probability of each base pair to be formed in a polynomial time frame \cite{mccaskill1990equilibrium}.

\subsection{Experimental data}

\subsubsection{Chemical probing data.}
Reactivities for systems \texttt{1KXK}, \texttt{2GIS}, \texttt{3IRW}, \texttt{3SD3}, \texttt{3VRS} and \texttt{4XW7} 
were collected for this work.
Single stranded DNA templates containing the T7 promoter region and the 3’ and 5’ SHAPE cassettes \cite{merino2006protocol} 
were ordered from Eurofins Genomics. RNAs were transcribed using in-house prepared T7 polymerase. 
Briefly, complementary T7 promoter DNA was mixed with the desired DNA template and snap cooled 
(\SI{95}{\celsius} for 5 minutes, followed by incubation on ice for 10 minutes) to ensure annealing of the T7 complementary 
promoter with template DNA. The mixture was supplemented with rNTPs, 20X transcription 
buffer (TRIS pH $8$, $\SI{100}{\milli\Molar}$ Spermidine, \SI{200}{\milli\Molar} DTT), PEG 8000, various concentrations of MgCl2 (final concentration ranging from $10$ to $\SI{40}{\milli\Molar}$), and T7 (\SI{10}{\milli\gram/\milli\liter}).
The RNAs were purified under denaturing conditions using polyacrylamide gel electrophoresis. RNA was excised from the gel and extracted using the crush and soak
method \cite{mori1993crushandsoak}. Following crush and soak, the RNAs were precipitated using ethanol and sodium acetate, and resuspended
in RNAse-free water. SHAPE modification followed by reverse transcription (using 5' FAM labeled primers) was carried out
as previously described \cite{merino2006protocol}. Following reverse transcription, RNAs were precipitated using ethanol and sodium acetate, 
redissolved in HiDi formamide, and cDNA fragments separated using capillary electrophoresis (ABI 3130 Sequencer). 
Raw reads corresponding to cDNA fragments were obtained using QuSHAPE \cite{karabiber2013qushape} and
are reported in Supporting Information.
Reads in each of the control and modifier channels were first normalized independently by dividing them
by the sum of reads in the corresponding channel. Reactivities were then estimated by subtracting
the normalized reads in the control channel from the normalized reads in the modifier channel, with negative
values replaced with zeros. This normalization is a simplified version of the one proposed in Ref.~\cite{aviran2011normalize} 
and does not contain position dependent corrections.
%These corrections are only expected to be relevant for RNA molecules significantly larger than those analyzed here.
Reactivities to different chemical probes, namely 1M7 for systems \texttt{1EHZ}, \texttt{1NBS}, \texttt{1Y26}, \texttt{2GDI}, \texttt{3DIG}, \texttt{3IGI},
\texttt{3PDR}, \texttt{4L81}, \texttt{4YBB\_CB} and \texttt{5KPY}, NMIA for \texttt{1NBS}, and DMS for \texttt{4P8Z} and \texttt{4QLM},
were taken from the literature \cite{cordero2012rna,loughrey2014shape,hajdin2013shape,poulsen2015shapes} and were normalized using the
procedure discussed above, except where noted.

\subsubsection{DCA data.}

Direct couplings for all the systems were calculated using the same code and parameters reported in Ref.~\cite{cuturello2018assessing},
but alignment was performed with ClustalW \cite{larkin2007clustal} to avoid including indirectly known structural information.
For systems where the RNA primary sequences used in Ref.~\cite{cuturello2018assessing} were different from those reported in the PDB or used in
chemical probing experiments, DCA calculations were performed again.
Couplings $J_{ij}$ were computed as the Frobenius norm of the couplings between positions $i$ and $j$,
as detailed in Ref.~\cite{cuturello2018assessing}.
All the used alignments and couplings are reported in Supporting Information.

\subsection{Penalties}

We integrate chemical probing reactivities $R_i$ and direct couplings $J_{ij}$ into the model by mapping them into single-point penalties $\lambda_i$ 
and pairwise penalties $\lambda_{ij}$ to pairing propensity of, respectively, individual nucleotides and specific nucleotide pairs. 
The free energy estimate $F\left(\vec{s} | \vec{seq}; \vec{R}, \vec{J}\right)$ obtained in this way is a modification of the original one by two additional terms:
$
F\left(\vec{s} | \vec{seq}; \vec{R}, \vec{J}\right)=F_0\left(\vec{s} | \vec{seq}\right)
+RT\sum\limits_{i=1}^{l_{seq}} \lambda_i(\vec{R})\cdot \left(1-s_i\right) + RT\sum\limits_{j>i+2}^{l_{seq}}\lambda_{ij}(\vec{J})\cdot s_{ij}
$
where $s_{i}$ is the pairing status of the $i$-th nucleotide in the structure $\vec{s}$ ($s_i=1$ if nucleotide $i$ is paired, $s_i=0$ otherwise)
%\begin{equation}
%  s_i = \left \{
%  \begin{aligned}
%    &1 && \text{if nucleotide i is paired}\\
%    &0 && \text{otherwise}
%  \end{aligned} \right.
%\end{equation}
and $s_{ij}$ is the pairing status of the specific couple of nucleotides $i$ and $j$.
%\begin{equation}
%  s_{ij} = \left \{
%  \begin{aligned}
%    &1 && \text{if nucleotide i is paired with nucleotide j} \\
%    &0 && \text{otherwise}
%  \end{aligned} \right.
%\end{equation}
We implement both kinds of penalties in the folding algorithm using the soft constraints functions 
from RNAlib \texttt{vrna\_sc\_add\_up} and \texttt{vrna\_sc\_add\_bp}, respectively.
We notice that penalties on individual nucleotides are used in several
methods developed to account for chemical probing experiments \cite{zarringhalam2012integrating,washietl2012rna}
though the way
these penalties are computed can differ.
Also notice that the most used model to include SHAPE data in secondary structure prediction
\cite{deigan2009accurate}
uses slightly different penalties that are associated to consecutive
rather than to individual base pairs.

\subsection{Neural network}

An important ingredient in our procedure is the way experimental data (reactivities and direct couplings)
are mapped into single and pairwise penalties, respectively.

The penalties associated with individual nucleotides are mapped from reactivities via a 
single-layered convolutional network
$
\lambda_i\left(\vec{R}\right)=\sum\limits_{k=-p}^p a_k\cdot R_{i+k} + b
$
that includes reactivities $\{R_{i-p},\dots ,R_{i-1}\}$ of the $p$ neighbor nucleotides in the $3'$ direction and reactivities
$\{R_{i+1},\dots ,R_{i+p}\}$ of the $p$ neighbor nucleotides in the $5'$ direction.
Hence, the hyperparameter $p$ determines the size of the convolutional window, namely $2p+1$. 
The parameters $a_k$ of the linear activation function control the relative weights of neighbors, and $b$ is the bias.

The penalties on specific nucleotide pairs are mapped from direct couplings via a double-layered network
$
\lambda_{ij}\left(J_{ij}\right)=C\cdot\sigma\left(A\cdot J_{ij}+B\right)+D
$.
The activation function of the output layer is linear with parameters $C$ and $D$, whereas we apply 
a sigmoid activation $\sigma\left(x\right)=\frac{1}{1+e^{-x}}$ at the innermost layer, with weight $A$ and bias $B$.

The model has thus $2p+6$ free parameters:
$\{a_k,b\}$ for the penalties associated to the chemical probing data and $\{A,B,C,D\}$ for those associated to the DCA data.

\subsection{Training}

The modifications to the model free energy affect the whole ensemble of structures for a given sequence, resulting in modified populations
$
P\left(\vec{s} | \vec{seq}, \vec{R}, \vec{J}\right)=e^{-\frac{1}{RT}F\left(\vec{s} | \vec{seq}; \vec{R}, \vec{J}\right)}/Z\left(\vec{seq}, \vec{R}, \vec{J}\right)
$.
Our aim is to increase the population of the native structure, under the assumption that the native structure is the one obtained by X-ray crystallography.
We thus consider a set of given sequence-structure pairs $\{\vec{seq},\hat{s}\}$ (one for each system in the training set),
where $\hat{s}$ denotes the pairing state in an available crystallographic structure,
and for each system we train the model to minimize the cost function
$
\mathcal{C}\left(\{a_k,b\}, \{A,B,C,D\}\right)=-RT\ln P\left(\hat{s} | \vec{seq}, \vec{R}, \vec{J}\right)
$.
 Its minimization, in the training procedure, is equivalent to maximizing the population of the target structures. 

For each system we decompose the cost function into two terms, namely $F\left(\vec{s} | \vec{seq}; \vec{R}, \vec{J}\right)$ 
and $-RT\ln Z\left(\vec{seq}, \vec{R}, \vec{J}\right)$ that we can compute using, respectively, 
the functions \texttt{vrna\_eval\_structure} and \texttt{vrna\_pf} from RNAlib. 
The derivatives of the cost function with respect to model paramaters, that are required for cost minimization, are proportional 
to pairing probabilities of individual nucleotides $p_i$ and of specific nucleotide pairs $p_{ij}$.
%\begin{equation}\label{derivatives}
%\begin{aligned}
%        &\frac{\partial \mathcal{C}}{\partial a_n}=RT\sum\limits_{i=1}^{l_{seq}}\left(p_i-\hat{s}_i\right)\frac{\partial\lambda_i}{\partial a_n} 
%        =RT\sum\limits_{i=1}^{l_{seq}}\left(p_i-\hat{s}_i\right)R_{i+n}\\
%        &\frac{\partial \mathcal{C}}{\partial b}=RT\sum\limits_{i=1}^{l_{seq}}\left(p_i-\hat{s}_i\right)\frac{\partial\lambda_i}{\partial b} 
%        =RT\sum\limits_{i=1}^{l_{seq}}\left(p_i-\hat{s}_i\right)\\
%        &\frac{\partial \mathcal{C}}{\partial A}=RT\sum\limits_{j>i+2}^{l_{seq}}\left(\hat{s}_{ij}-p_{ij}\right)\frac{\partial\lambda_{ij}}{\partial A} 
%        =RT\sum\limits_{j>i+2}^{l_{seq}}\left(\hat{s}_{ij}-p_{ij}\right)J_{ij}\cdot C\sigma'\left(AJ_{ij}+B\right)\\
%        &\frac{\partial \mathcal{C}}{\partial B}=RT\sum\limits_{j>i+2}^{l_{seq}}\left(\hat{s}_{ij}-p_{ij}\right)\frac{\partial\lambda_{ij}}{\partial B} 
%        =RT\sum\limits_{j>i+2}^{l_{seq}}\left(\hat{s}_{ij}-p_{ij}\right)\cdot C\sigma'\left(AJ_{ij}+B\right)\\
%        &\frac{\partial \mathcal{C}}{\partial C}=RT\sum\limits_{j>i+2}^{l_{seq}}\left(\hat{s}_{ij}-p_{ij}\right)\frac{\partial\lambda_{ij}}{\partial C} 
%        =RT\sum\limits_{j>i+2}^{l_{seq}}\left(\hat{s}_{ij}-p_{ij}\right)\sigma\left(AJ_{ij}+B\right)\\
%        &\frac{\partial \mathcal{C}}{\partial D}=RT\sum\limits_{j>i+2}^{l_{seq}}\left(\hat{s}_{ij}-p_{ij}\right)\frac{\partial\lambda_{ij}}{\partial D} 
%        =RT\sum\limits_{j>i+2}^{l_{seq}}\left(\hat{s}_{ij}-p_{ij}\right)\\
%\end{aligned}
%\end{equation}
These derivatives are then used to back propagate derivatives from the output layer to the input nodes.
%Here $\hat{s}_i$ and $\hat{s}_{ij}$ represent the pairing state in the reference structure for nucleotide $i$ and for pair $ij$ respectively.
Base-pair probabilities in the penalty-driven ensembles $p_{ij}=\sum_{\{\vec{s}\}}P\left(\vec{s} | \vec{seq}, \vec{R}, \vec{J}\right)s_{ij}$,
can be straightforwardly computed using the function \texttt{vrna\_bpp} from RNAlib.
  
\subsection{Regularization}

In order to reduce the risk of overfitting we include $\textit{l}-2$ regularization in the training procedure. Direct couplings (two-dimensional data)
and reactivity profiles (one-dimensional data) differ in the amount of structural information they contain. For this reason, instead of adding to the cost function
a standard single regularization term on all parameters, we add two representational regularization
terms \cite{goodfellow2016deep}, each with an independent coefficient, directly on the penalties mapped from
each type of data,
$
\mathcal{C}\left(\{a_k,b\}, \{A,B,C,D\}\right)=-RT\ln P\left(\hat{s} | \vec{seq}, \vec{R}, \vec{J}\right)
+\alpha_S\sum_{i}\lambda_i^2+\alpha_D\sum_{ij}\lambda^2_{ij}
$.
This procedure keeps the penalties that we add to the model free energy from becoming too large, and thus helps preventing 
the occurence of overfitting during the minimization of the cost function.
The introduction of regularization terms must be taken into account in the cost function derivatives
by addition of corresponding derivative terms that are easily computed.

\subsection{Minimization}

The inclusion of regularization terms in the cost function brings in two hyperparameters, $\alpha_S$ and $\alpha_D$,
in addition to $p$, the hyperparameter that determines the width of the convolutional window.
The collection of models that we train is thus defined by the triplet of hyperparameters $\{p,\alpha_S,\alpha_D\}$.
We then explore all the hyperparameter combinations within the ranges
$p\in [0,1,2,3]$ and 
$\alpha_S,\alpha_D\in[\infty,1.0,10^{-1},10^{-2},10^{-3},10^{-4}, 0.0]$
for a total of $4\times7\times7=196$ models.
For each model, we minimize the corresponding cost function using the sequential quadratic programming algorithm
as implemented in the \texttt{scipy.optimize} optimization package \cite{virtanen2020scipy}.
The minimization problem is non-convex whenever $\alpha_D$ is finite, so we expect the cost function landscape to be rough,
with multiple local minima. The result of the minimization will thus depend on the initial set of model parameters.
For each minimization we try multiple initial values for the model parameters,
extracting them from a random uniform distribution, and we select those that yield the minimum cost function.
For each minimization we include in the set of starting parameters also three 
specific sets of starting points:
\begin{itemize}
\item parameter values from the optimized $\{p-1,\alpha_S,\alpha_D\}$ model, with the new $a_{-p}$ and $a_p$ set to $0.0$;
 if $p=0$, we ignore this starting point.
\item parameter values from the optimized $\{p,10\cdot\alpha_S,\alpha_D\}$ model;
      if $\alpha_S=0.0$, we use values from the optimized $\{p,10^{-4},\alpha_D\}$ model;
      if $\alpha_S=1$, we use values from the optimized $\{p,\infty,\alpha_D\}$ model;
      if $\alpha_S=\infty$, we ignore this starting point.
\item parameter values from the optimized $\{p,\alpha_S,10\cdot\alpha_D\}$ model;
      if $\alpha_D=0.0$, we use values from the optimized $\{p,\alpha_S,10^{-4}\}$ model;
      if $\alpha_D=1$, we use values from the optimized $\{p,\alpha_S,\infty\}$ model;
      if $\alpha_D=\infty$, we ignore this starting point.
\end{itemize}
This ensures that models with higher complexity (\textit{i.e.}, higher $p$ or lower $\alpha_S$ or $\alpha_D$) will, by construction, fit the data better than models with lower complexity.
In this way the performance of the models, as evaluated on the training set, is by construction a monotonically decreasing function
of $\alpha_D$ and $\alpha_S$, and a monotonically increasing function of $p$.

\subsection{Leave-one-out}

Among the models optimized in the training procedure, we select the one that yields the best performance without 
overfitting the training data, in order to ensure the transferability of its structure and optimal parameters.
As a test for transferability, we use a leave-one-out cross-validation. This procedure consists in iteratively
leaving each of the 12 systems at a time out of the training set, and using the optimal parameters 
resulting from optimization on the reduced training set to compute the population of the native structure 
for the left-out system. The population of native structures, averaged on the left-out systems, is used to rank
all of the tested models. We consider the model with the highest score as the most capable of yielding an increase 
in population of native structures for systems on which it was not trained.

\subsection{Validation}

The resulting model is then validated on a set of 6 systems that were not used in the parameter or hyperparameter optimization.
For these systems we compute the ensemble population of the native structure. In addition, we compute the similarity between the
most stable structure in the predicted ensemble (minimum free energy structure) and the native structure using the Matthews correlation
coefficient, that optimally balances sensitivity and precision.

\section{RESULTS}

Chemical probing experiments provide reactivities per nucleotide (one-dimensional information, $R_i$) that are mapped via a single-layered convolutional network to penalties to be associated to the pairing propensity of individual nucleotides ($\lambda_i$).
Similarly, direct-coupling analysis provides predicted contact scores (two-dimensional information, $J_{ij}$)
that are mapped through a non-linear function into penalties to be associated with specific nucleotide-nucleotide pairs ($\lambda_{ij}$).
The resulting penalties are integrated in the folding algorithm RNAfold from the Vienna package \cite{lorenz2011viennarna}, which allows the full partition function of the system to be computed,
including the population of any suboptimal structure. The parameters of the mapping functions are trained in order to maximise the
population of the secondary structures as annotated in a set of high-resolution X-ray diffraction experiments.
The differentiability of the RNAfold model with respect to the applied penalties is crucial, since it allows
the thermodynamic model to be used during the training procedure.
Reference structures are obtained from the structural database \cite{burley2018rcsb}. Reference chemical probing data are partly taken from the 
RNA mapping database \cite{cordero2012rna,loughrey2014shape} and from Refs.~\cite{hajdin2013shape,poulsen2015shapes}, 
and partly reported for the first time in this paper. Reference direct couplings are partly taken from Ref.~\cite{cuturello2018assessing} and partly
obtained in this paper, using RNA families deposited on RFAM \cite{nawrocki2014rfam}.
The model complexity is controlled via three hyperparameters, which are chosen using a cross-validation procedure, and the obtained model is evaluated on an independent dataset
not seen during the training procedure.
A more detailed explanation can be found in Materials and Methods, and the architecture of the model is summarized in Fig.~\ref{netscheme}.

\subsection{Model training}

\begin{figure*}
\centering
    \begin{subfigure}[t]{0.42\textwidth}
        \includegraphics[width=1.0\linewidth, height=7cm, trim={=0.0cm 2.5cm 0.0cm 4.5cm}, clip]{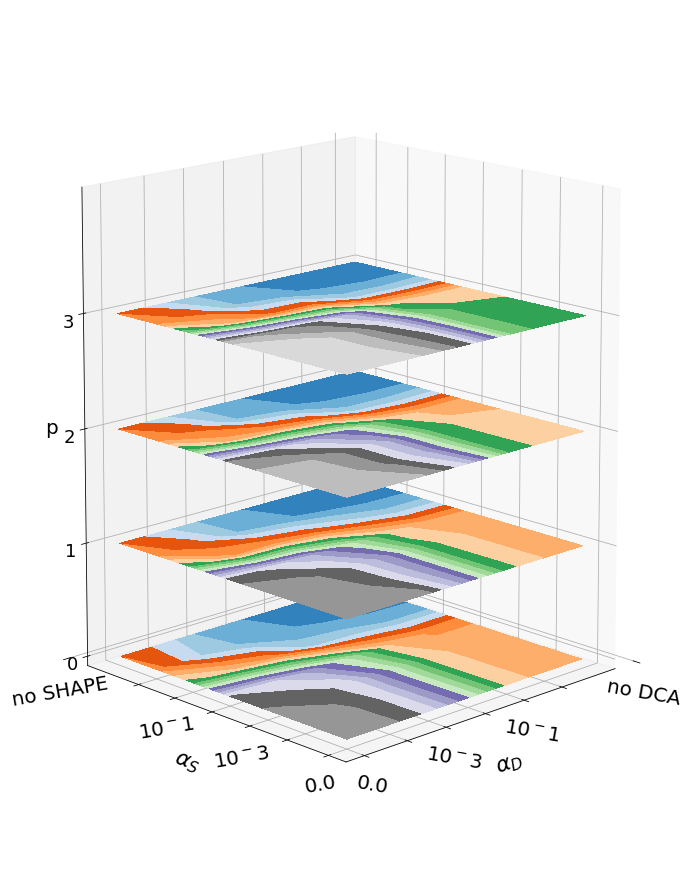}
        \caption{}
    \end{subfigure}\hfill
    \begin{subfigure}[t]{0.42\textwidth}
        \includegraphics[width=1.0\linewidth, height=7cm, trim={=0.0cm 2.5cm 0.0cm 4.5cm}, clip]{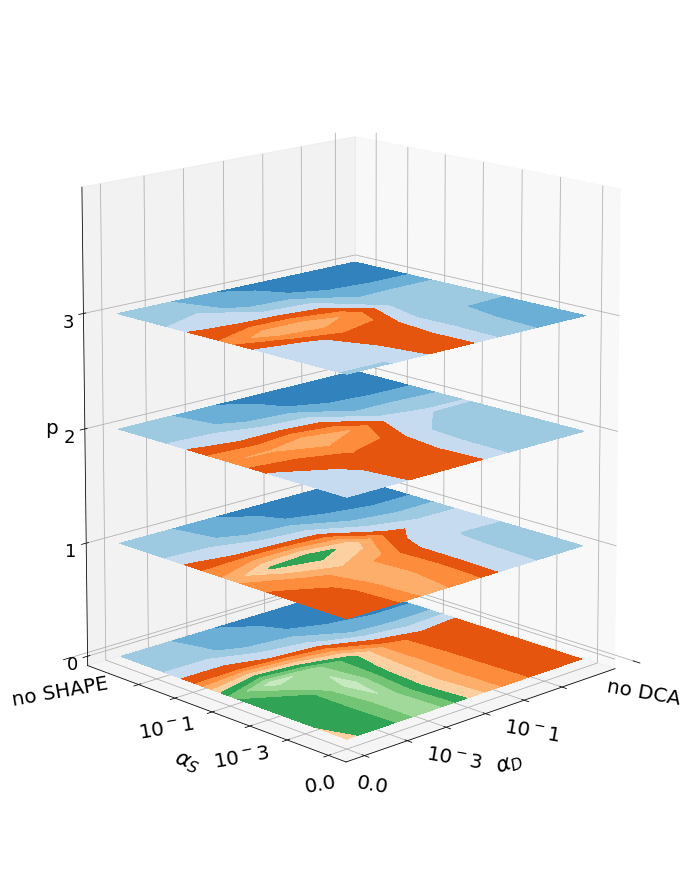}
        \caption{}
    \end{subfigure}\hfill
     \begin{subfigure}[t]{0.15\textwidth}
        \includegraphics[width=1.0\linewidth, height=7cm, keepaspectratio]{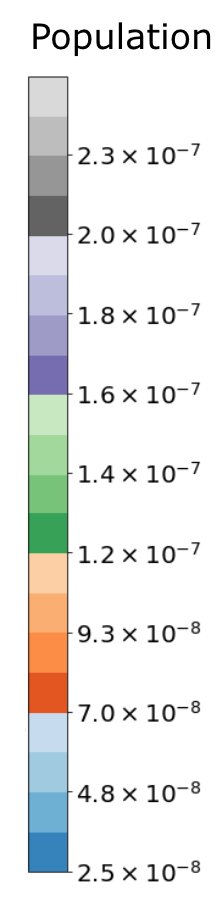}
    \end{subfigure}
    \caption{Population of native structure as function of hyperparameters.
Population is indicated in the color scale.
             The optimized population of native structures, when averaged on the training set (a), is by construction
             a monotonically increasing function of the integer $p$ controlling the window size of 
             the convolutional network in the reactivity channel, and a monotonically decreasing function
             of the regularization coefficients $\alpha_S$ and $\alpha_D$. When averaged
             on the leave-one-out iterations of the cross-validation (CV) procedure (b), the dependency of the optimized
             population of native structures on these hyperparameters becomes non-trivial, as it results
             from a combination of model complexity (controlled by $p$) and regularization
             (controlled by $\alpha_S$ and $\alpha_D$ independently). The CV procedure serves
             as criterion for model selection, resulting in the selection of hyperparameters
             $\{p=0,\alpha_S=0.001,\alpha_D=0.001\}$.
}
    \label{stackplot}
\end{figure*}

\begin{figure*}
\centering
    \begin{subfigure}[t]{0.49\textwidth}
        \includegraphics[width=1\linewidth]{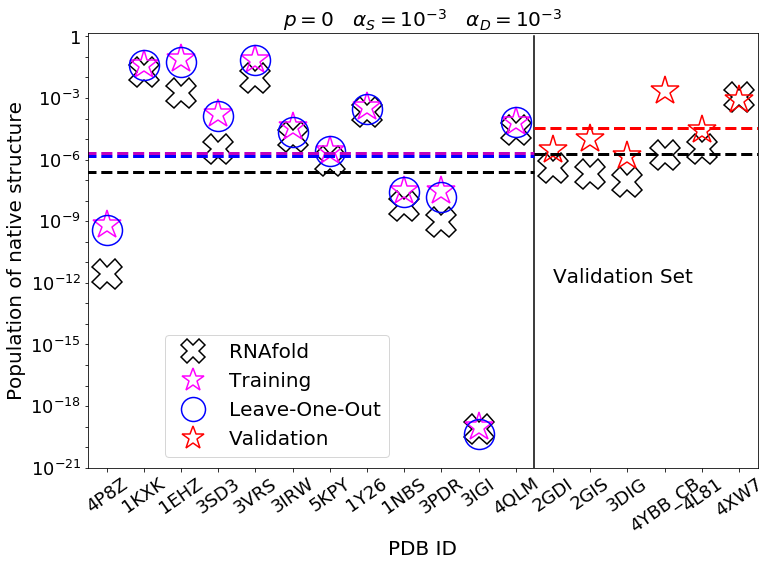}
        \caption{}
    \end{subfigure}\hfill
    \begin{subfigure}[t]{0.49\textwidth}
        \includegraphics[width=1\linewidth]{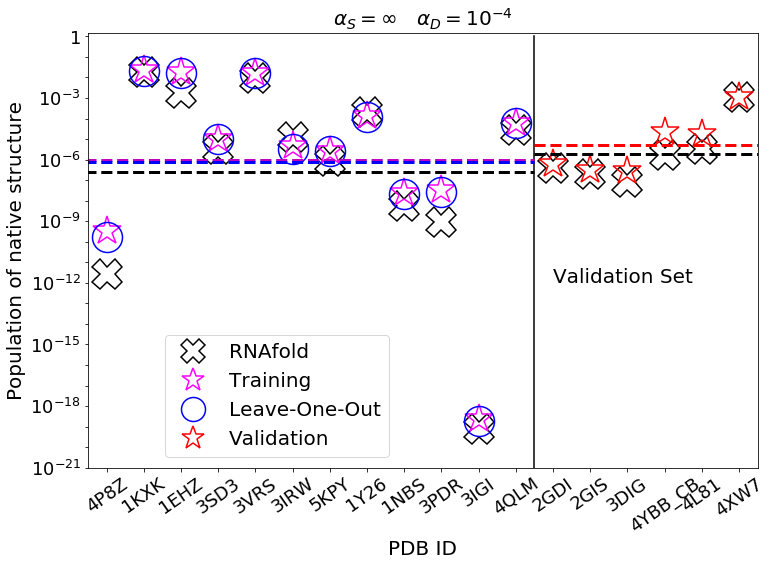}
        \caption{}
    \end{subfigure}\vfill
    \begin{subfigure}[t]{0.49\textwidth}
        \includegraphics[width=1\linewidth]{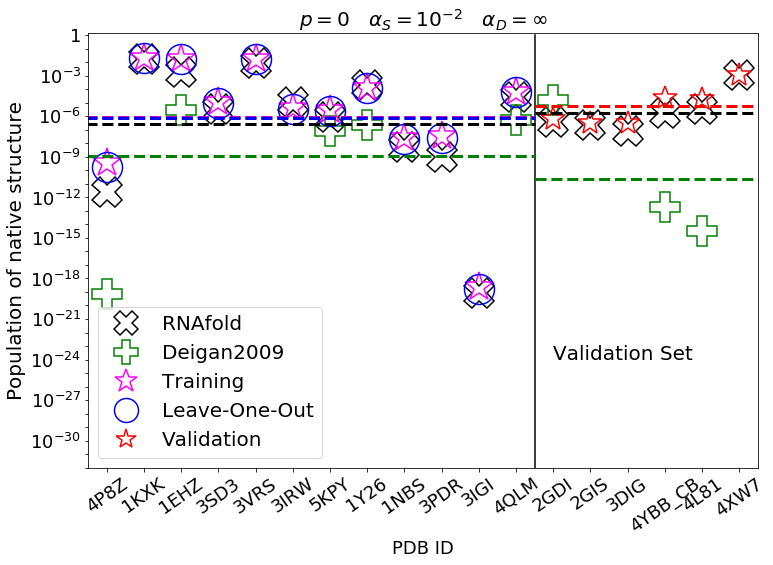}
        \caption{}
    \end{subfigure}\hfill
    \begin{subfigure}[t]{0.49\textwidth}
        \includegraphics[width=1\linewidth]{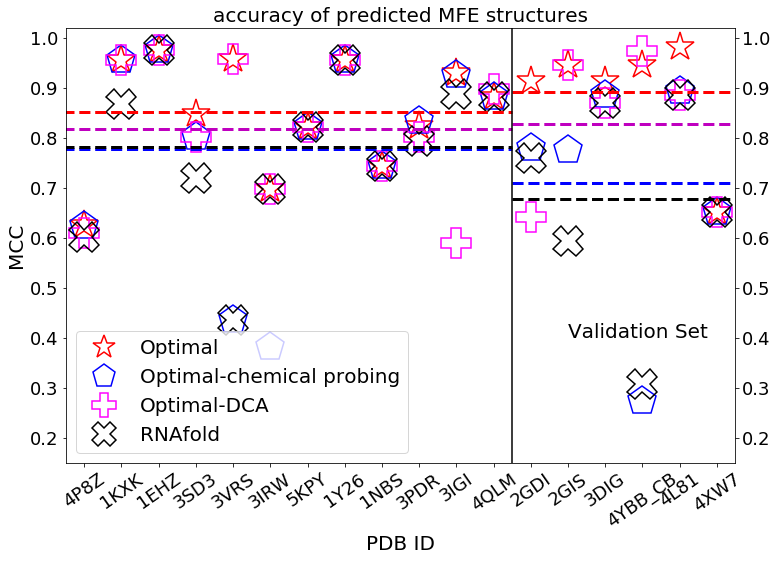}
        \caption{}
    \end{subfigure}
    \caption{Comparison of results obtained with unmodified RNAfold
     and with selected models, respectively: populations of native structures with (a) the best performing model; (b) the best performing model
     with DCA data only; (c) the best performing model with chemical probing data only. (d) Matthews correlation coefficients
     between predicted MFE structures and reference native structures, as obtained with selected (best, DCA-only, chemical probing-only)
     models and with unmodified RNAfold.
     Hyperparameters are noted in the figure.
     Native structure populations obtained with unmodified 
     RNAfold (black cross), with our trained model (magenta star on the training set, red star on the validation set) 
     and in the leave-one-out procedure
     (blue circle, for each molecule the model is trained on all the other molecules in the training set)
     are reported.
     Populations obtained by mapping SHAPE reactivities into penalties with the method in Ref.~\cite{deigan2009accurate} are reported for comparison (green plus),
     only for molecules studied in previous work and in panel (c) where chemical probing data only are used.
     The populations of native structures that we obtain with the trained model 
     are almost always increased for molecules both in the training (left side of the vertical line) and in the validation set (right side), 
     with overfitting occurring in a few cases, 
     where populations lower than obtained with unmodified RNAfold are yielded.     
     }
    \label{populations}
\end{figure*}
We randomly choose a training set of 12 systems, 
leaving 6 others out for later validation.
Since crystal structures, chemical probing data, and DCA data for different systems might be of different quality,
the specific choice of the splitting might affect the overall training and validation results.
We thus generate four independent random splittings, reported in Table \ref{dataset}.
In the following we refer to splitting S3, that leads to the worst performance in the cross-validation test and to the best performance 
in the external validation. Results for all the splittings are reported in Supporting Information. Importantly, the external validation 
test is passed for all the splittings, indicating that our procedure is capable to detect overfitting with all of the tested datasets.
The model complexity is controlled by means of three handles: a regularization parameter acting on the one-dimensional penalties
derived from reactivities ($0\le \alpha_S \le \infty$), a regularization parameter acting on the two-dimensional penalties
derived from DCA ($0\le \alpha_D \le \infty$) and an integer controlling the size of the window used for the convolutional network ($p<=3$).
When the performance of the model is evaluated on the training set, the model that better fits the data is
the most complex one, with no regularization term ($\alpha_S=\alpha_D=0$) and the largest tested window ($p=3$)
(Fig.~\ref{stackplot}a).
The geometric average of the populations of native structures increases by $\approx 11$ times 
with respect to that of the thermodynamic model alone.
Training the model using only chemical probing data
($\alpha_D=\infty$), or only DCA data ($\alpha_S=\infty$), results in an increase of native population
by $\approx$ 5 times and $\approx$ 3 times respectively, within the randomized set S3 (Table \ref{dataset}).

\subsection{Model selection}

In order to make the parametrization transferable, we perform a leave-one-out cross-validation (CV) procedure (see Materials and Methods)
where one of the 12 systems at a time is left out of the training procedure and the increase in the native population
for the left-out system is used to estimate transferability.
Overall, the average performance of the model on the left-out system shows a non-trivial dependence on the hyperparameters (Fig.~\ref{stackplot}b).
All the models yield a performance in the cross-validation test equal or better than the thermodynamic model alone, but
the best performance is obtained when choosing $\alpha_S=0.001$, $\alpha_D=0.001$ and $p=0$.
We select this model as the one that yields the best balance between performance and transferability.
Results obtained by using different randomizations of the training set are reported in Supporting Information.
Whereas the precise set of optimal hyperparameters depends on the specific training set, 
sets of hyperparameters that perform well on a specific set tend to perform well for all of the tested training sets.

\subsection{Validation on an independent dataset}

Finally, we evaluate the performance of the selected model on a dataset of 6 systems that were not seen during training.
This additional test is done in the spirit of nested cross-validation \cite{cawley2010over} in order to properly evaluate the
transferability of the procedure. 

\begin{figure*}
\centering
    \begin{subfigure}[t]{0.33\textwidth}
        \includegraphics[width=1\linewidth,height=3.5cm, keepaspectratio]{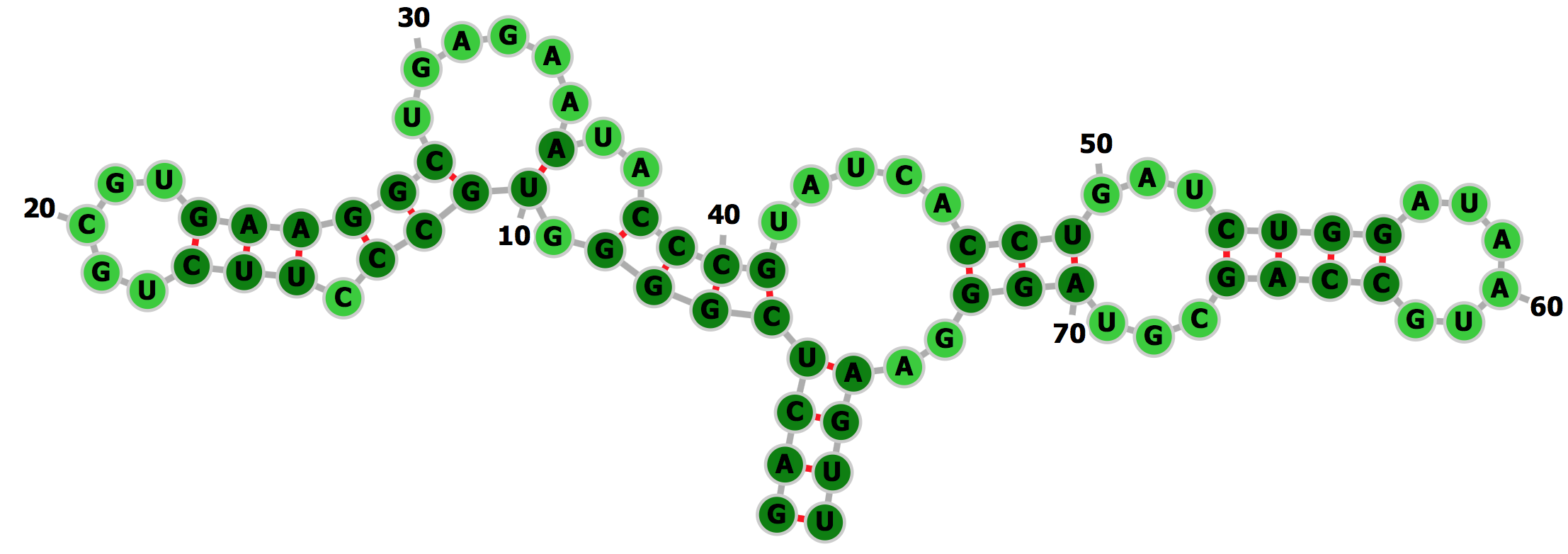}
        \caption{Reference}
    \end{subfigure}\hfill
    \begin{subfigure}[t]{0.33\textwidth}
        \includegraphics[width=1\linewidth,height=3.5cm, keepaspectratio]{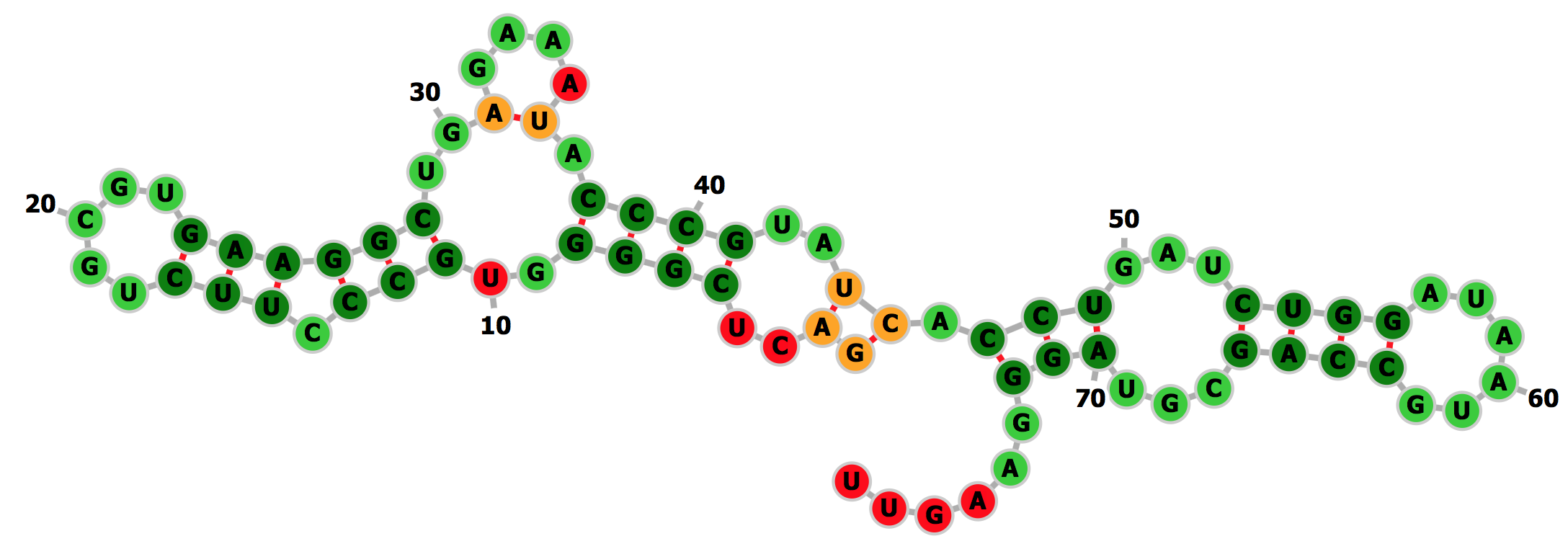}
        \caption{RNAfold (0.76)}
    \end{subfigure}\hfill
    \begin{subfigure}[t]{0.33\textwidth}
        \includegraphics[width=1\linewidth,height=3.5cm, keepaspectratio]{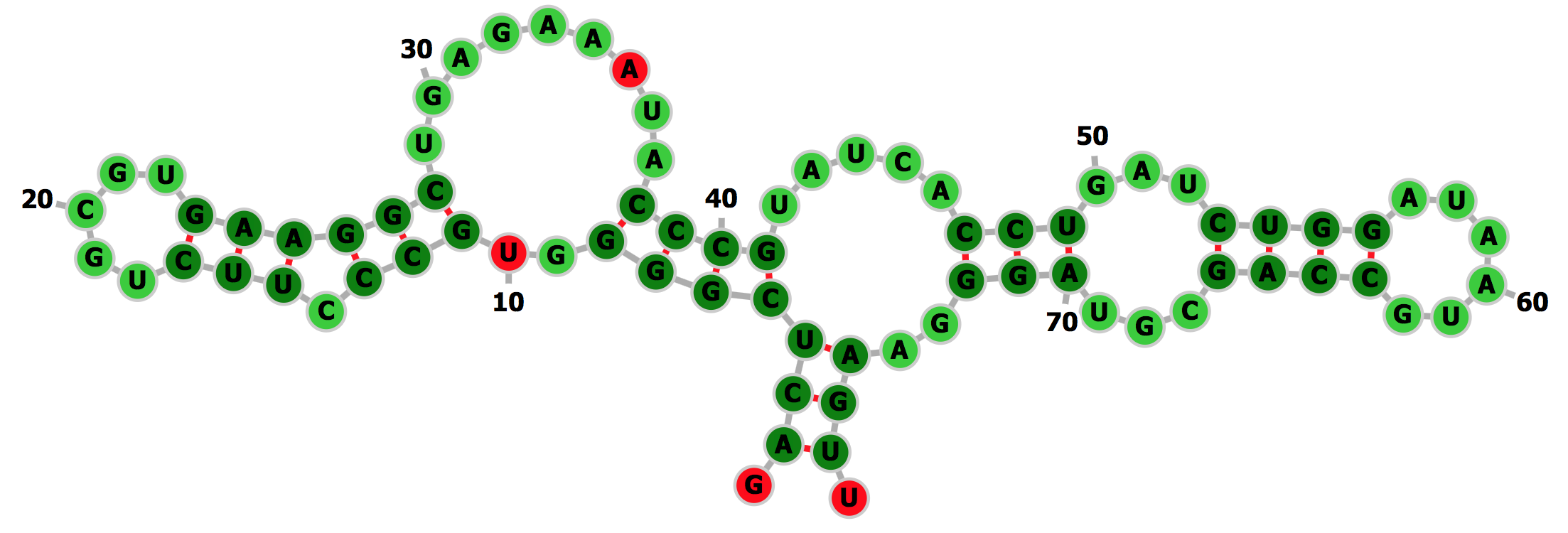}
        \caption{Trained model (0.91)}
    \end{subfigure}\vfill
    \begin{subfigure}[t]{0.33\textwidth}
        \includegraphics[width=1\linewidth,height=3.5cm, keepaspectratio]{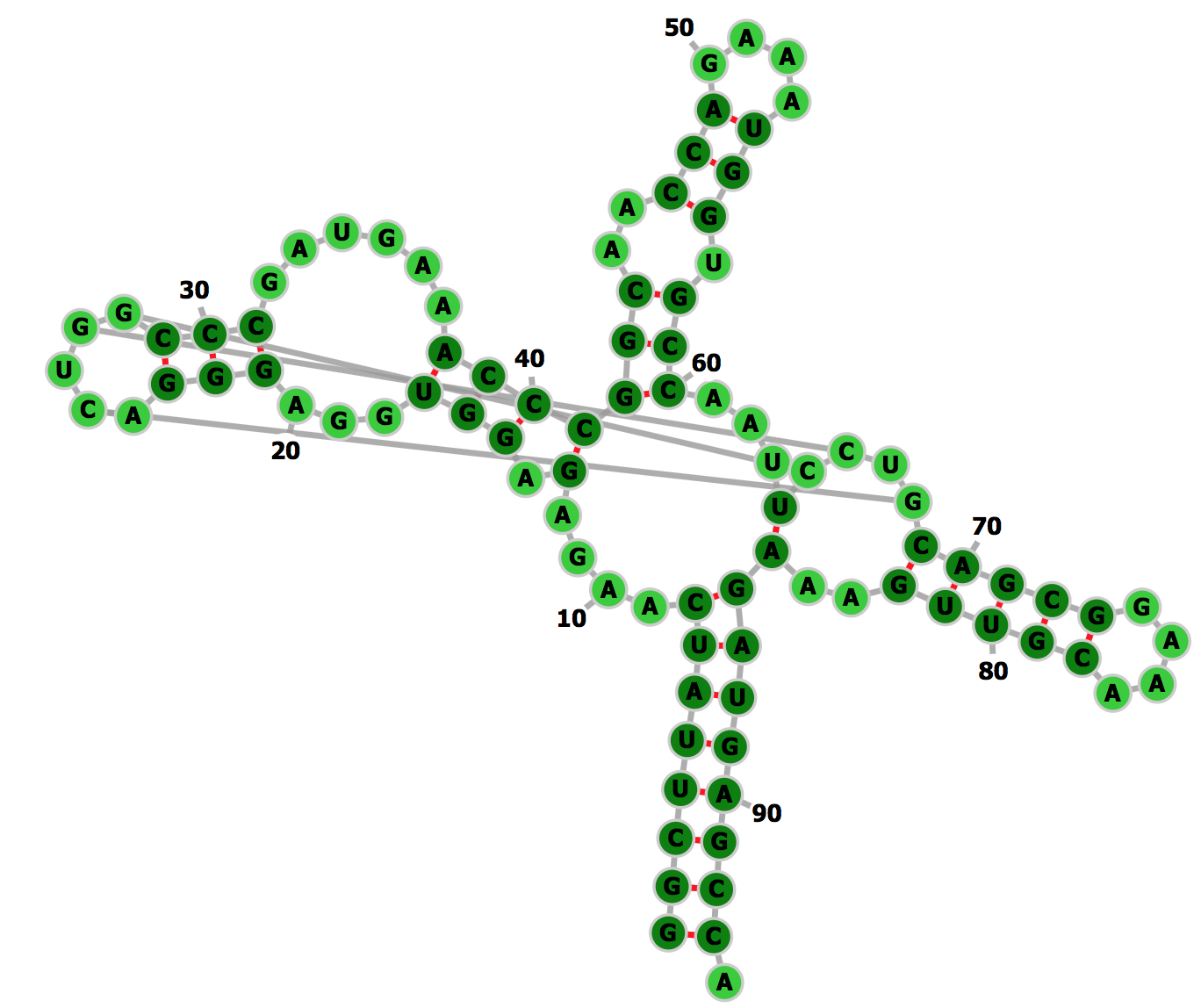}
        \caption{Reference}
    \end{subfigure}\hfill
    \begin{subfigure}[t]{0.33\textwidth}
        \includegraphics[width=1\linewidth,height=3.5cm, keepaspectratio]{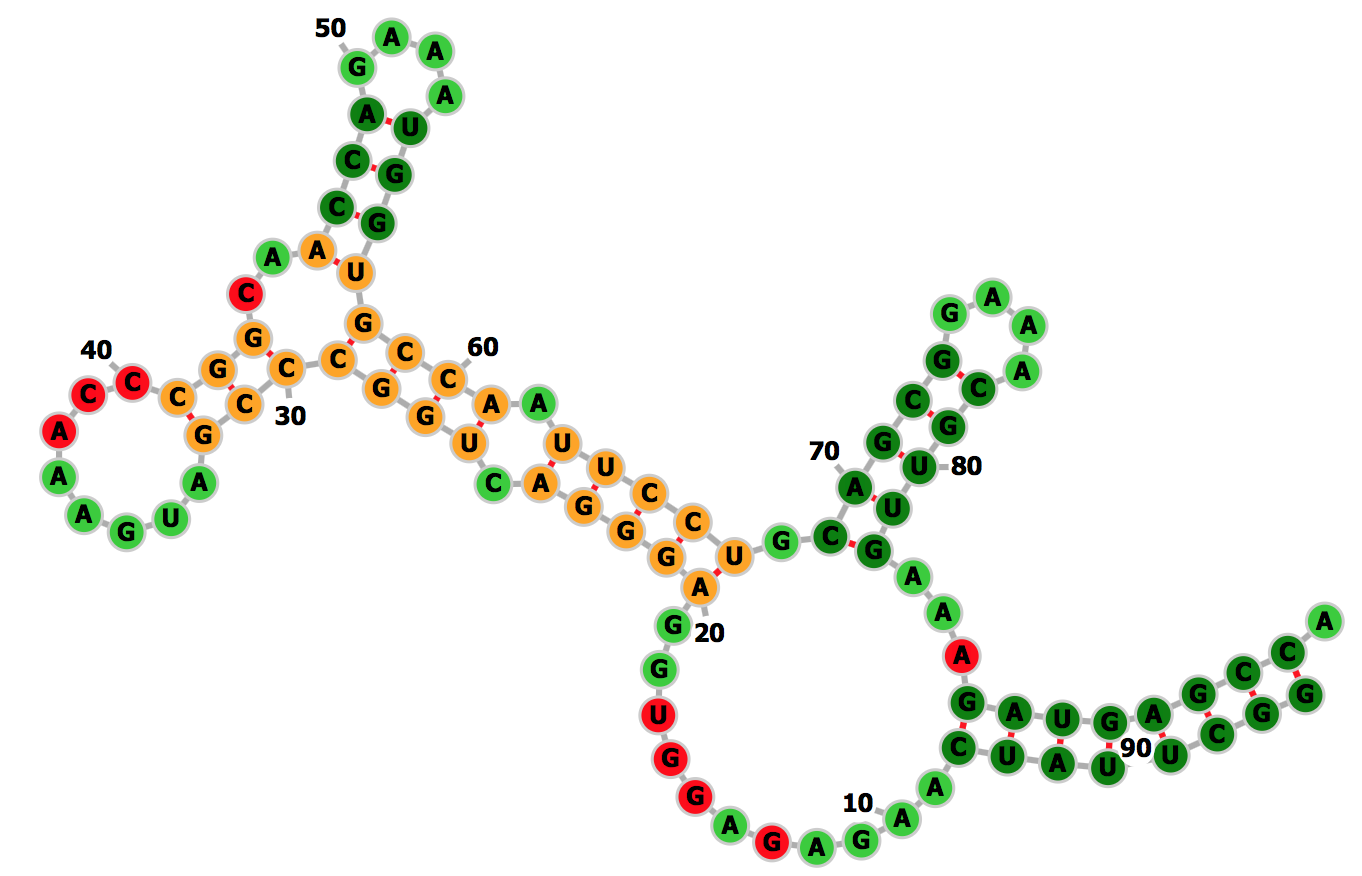}
        \caption{RNAfold (0.59)}
    \end{subfigure}\hfill
    \begin{subfigure}[t]{0.33\textwidth}
        \includegraphics[width=1\linewidth,height=3.5cm, keepaspectratio]{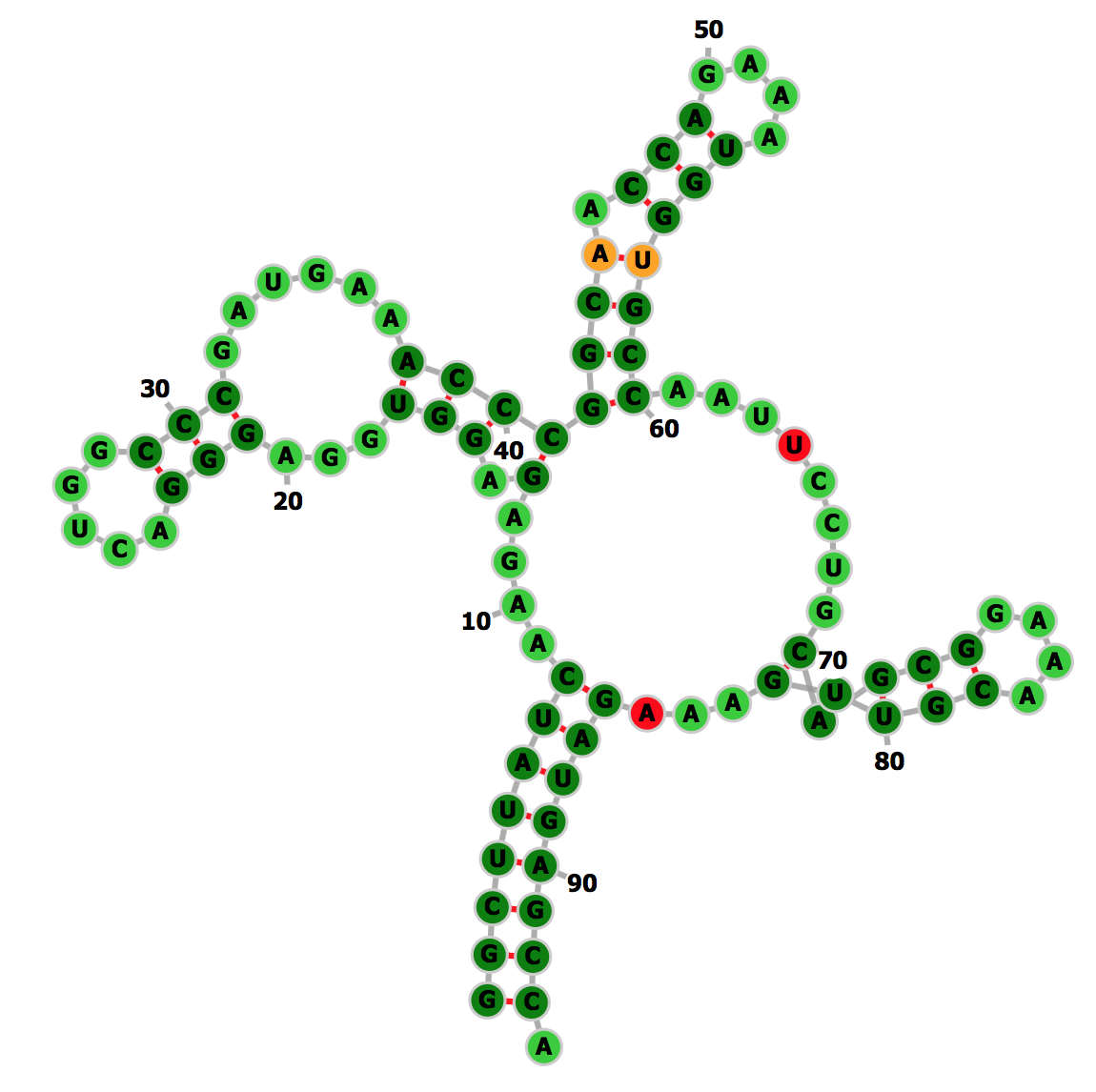}
        \caption{Trained model (0.95)}
    \end{subfigure}\vfill
    \begin{subfigure}[t]{0.33\textwidth}
        \includegraphics[width=1\linewidth,height=3.5cm, keepaspectratio]{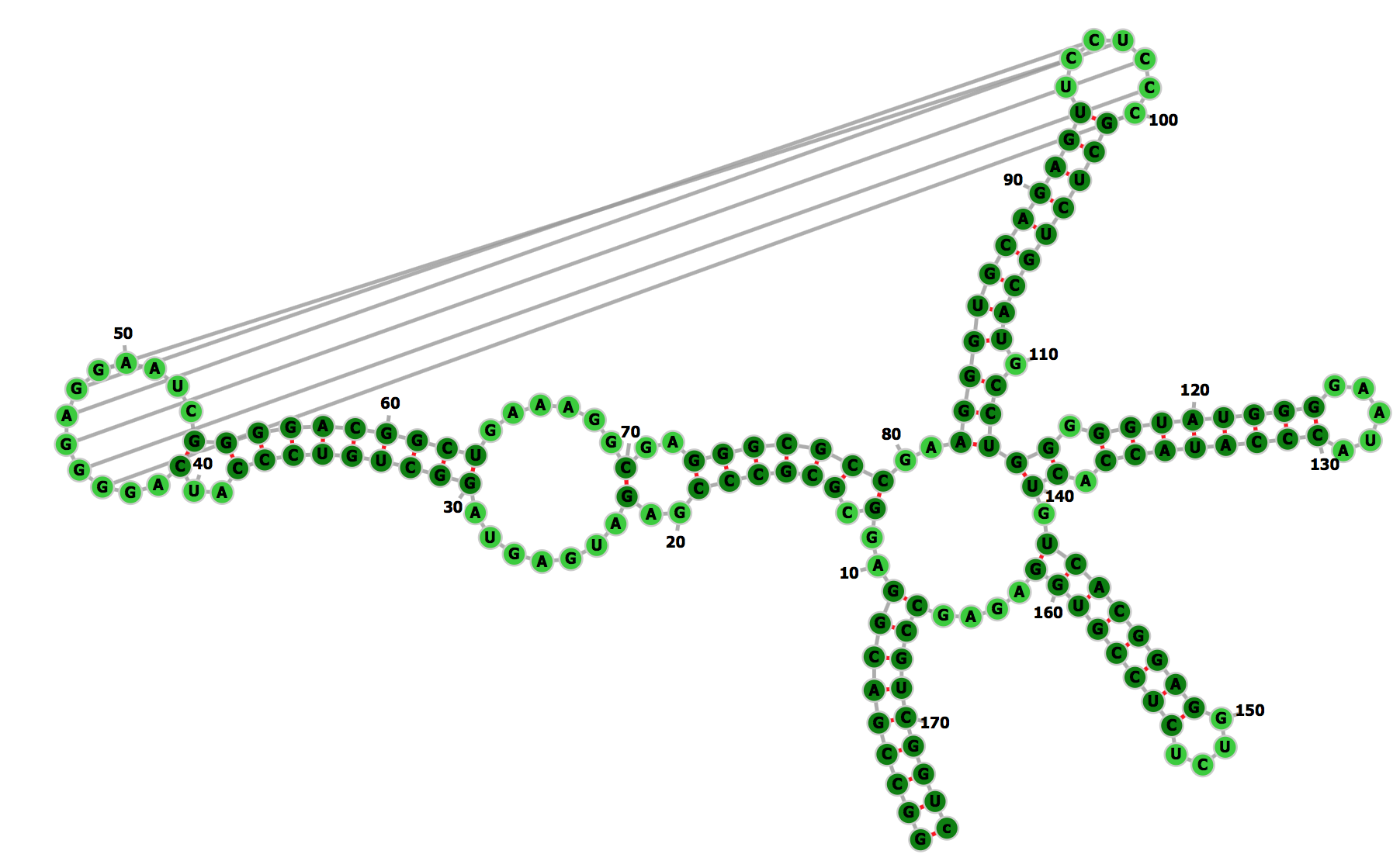}
        \caption{Reference}
    \end{subfigure}\hfill
    \begin{subfigure}[t]{0.33\textwidth}
        \includegraphics[width=1\linewidth,height=3.5cm, keepaspectratio]{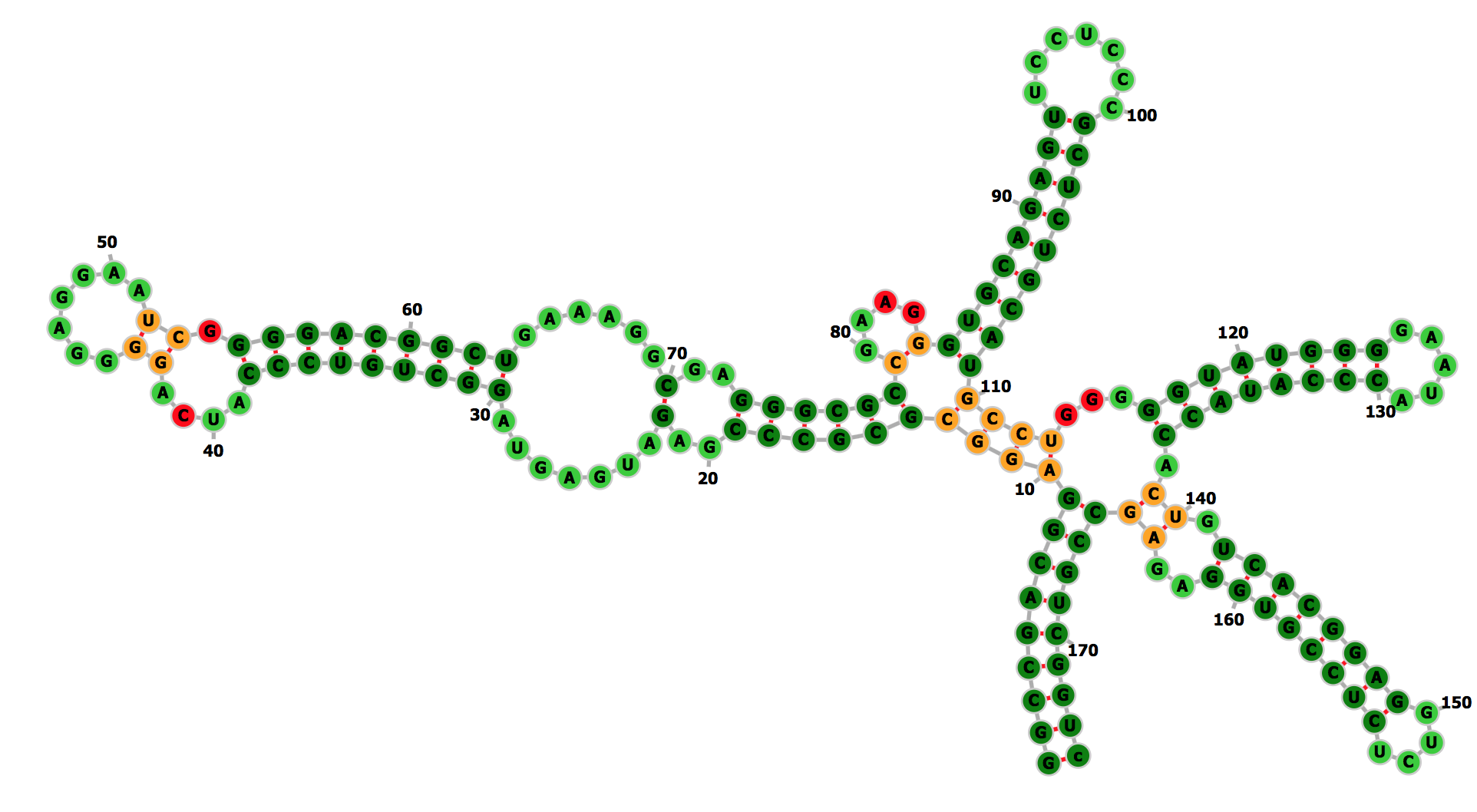}
        \caption{RNAfold (0.87)}
    \end{subfigure}\hfill
     \begin{subfigure}[t]{0.33\textwidth}
        \includegraphics[width=1\linewidth,height=3.5cm, keepaspectratio]{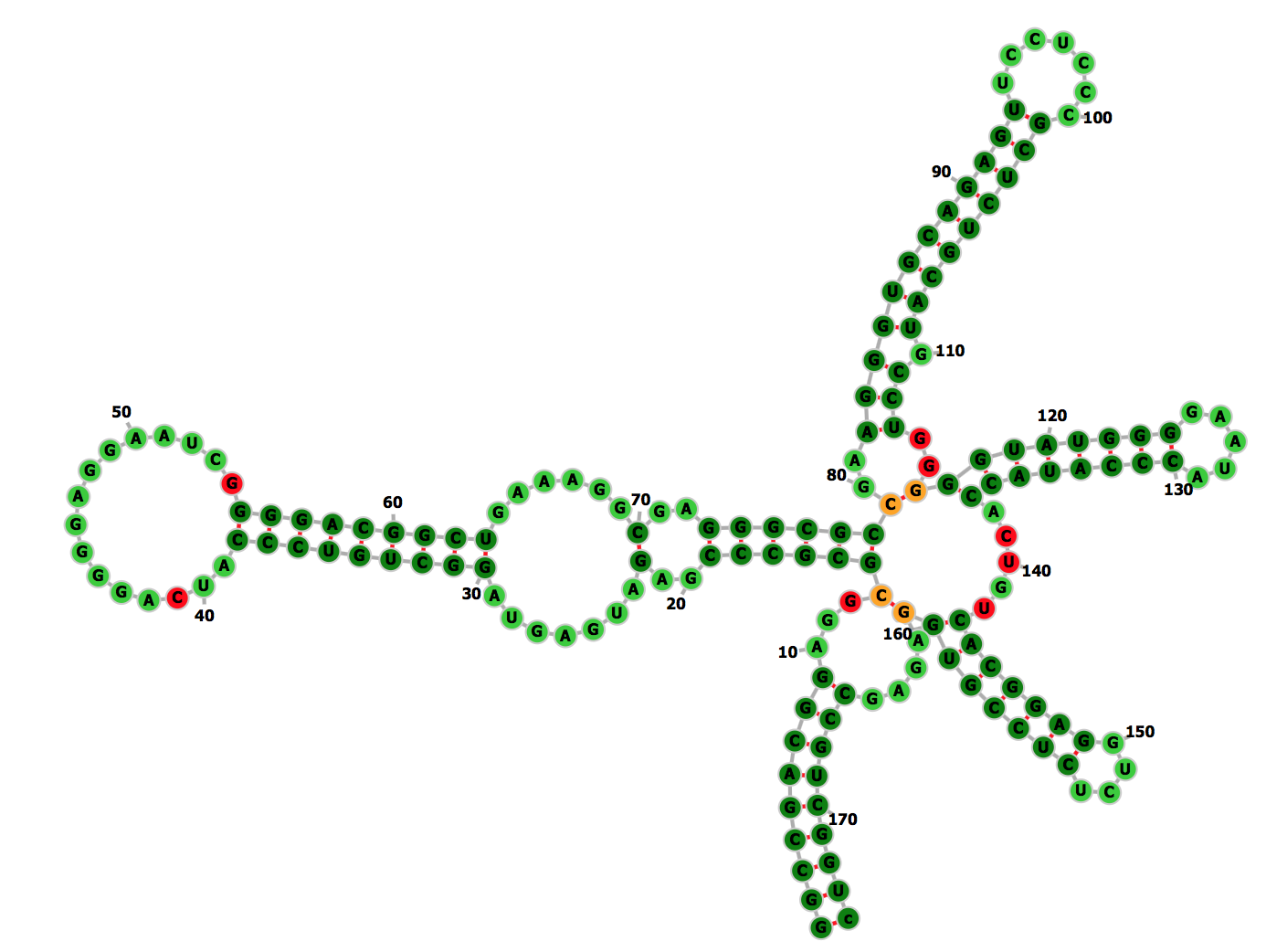}
        \caption{Trained model (0.91)}
    \end{subfigure}\vfill
     \begin{subfigure}[t]{0.33\textwidth}
        \includegraphics[width=1\linewidth,height=3.5cm, keepaspectratio]{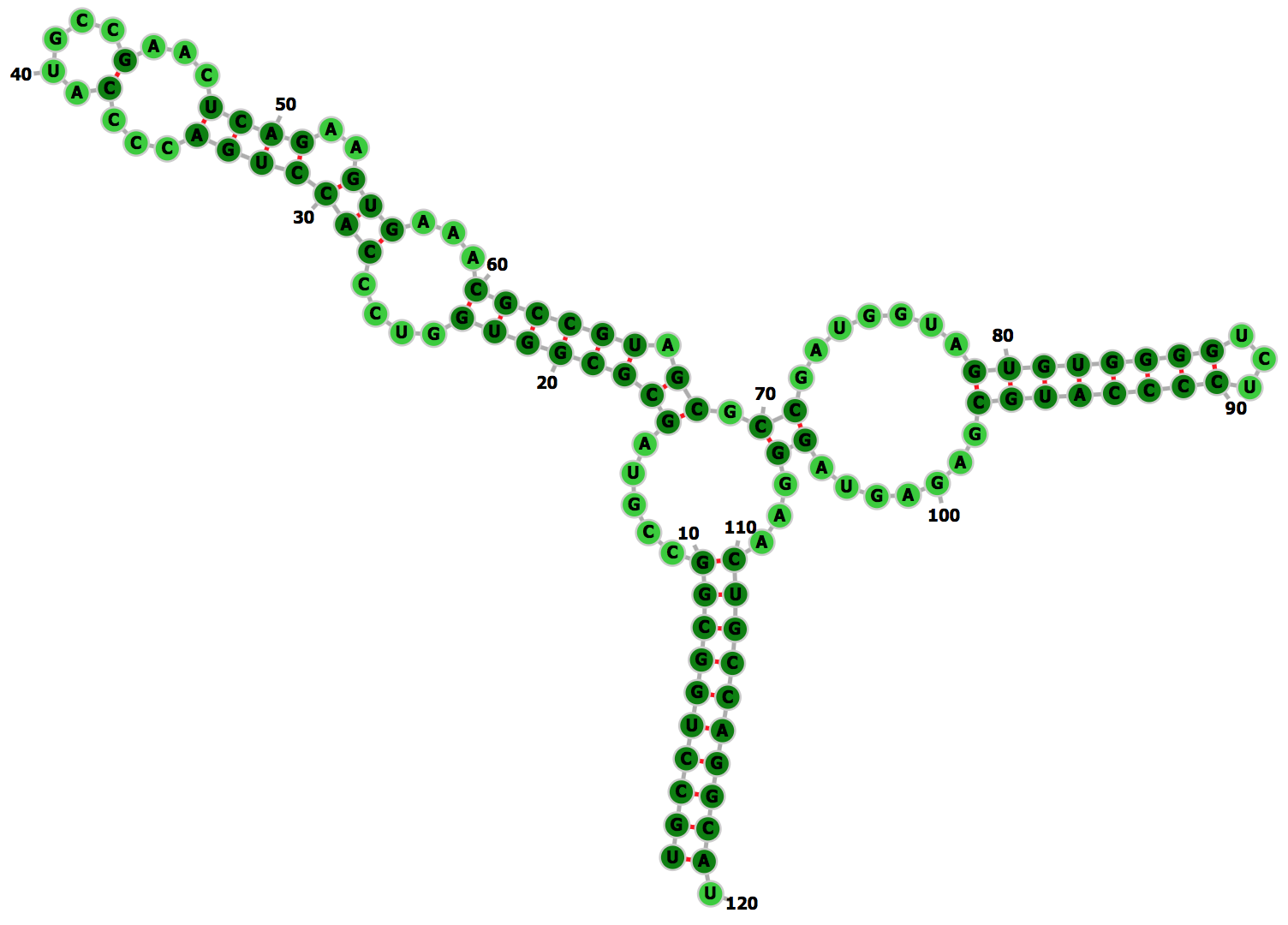}
        \caption{Reference}
    \end{subfigure}\hfill
    \begin{subfigure}[t]{0.33\textwidth}
        \includegraphics[width=1\linewidth,height=3.5cm, keepaspectratio]{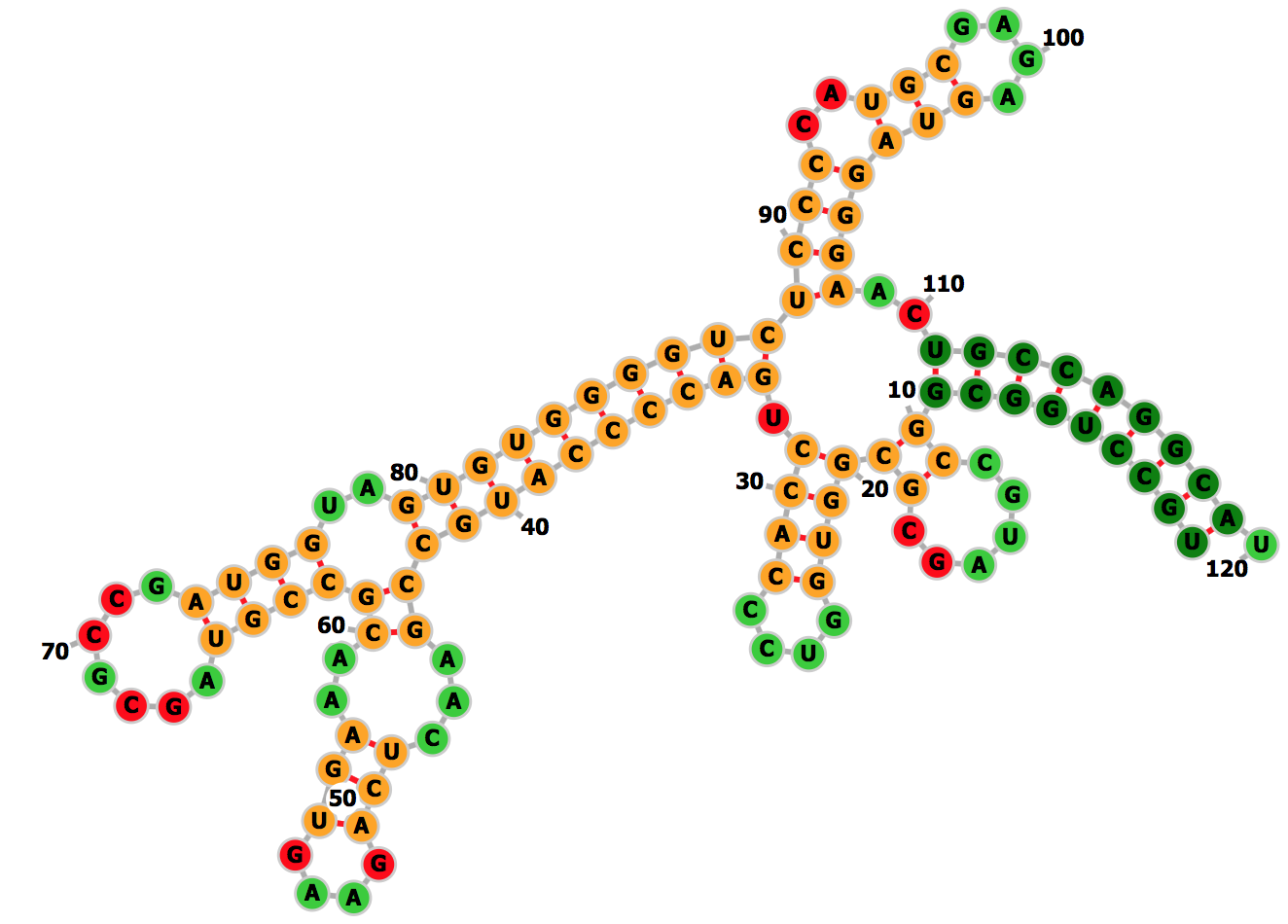}
        \caption{RNAfold (0.31)}
    \end{subfigure}\hfill
     \begin{subfigure}[t]{0.33\textwidth}
        \includegraphics[width=1\linewidth,height=3.5cm, keepaspectratio]{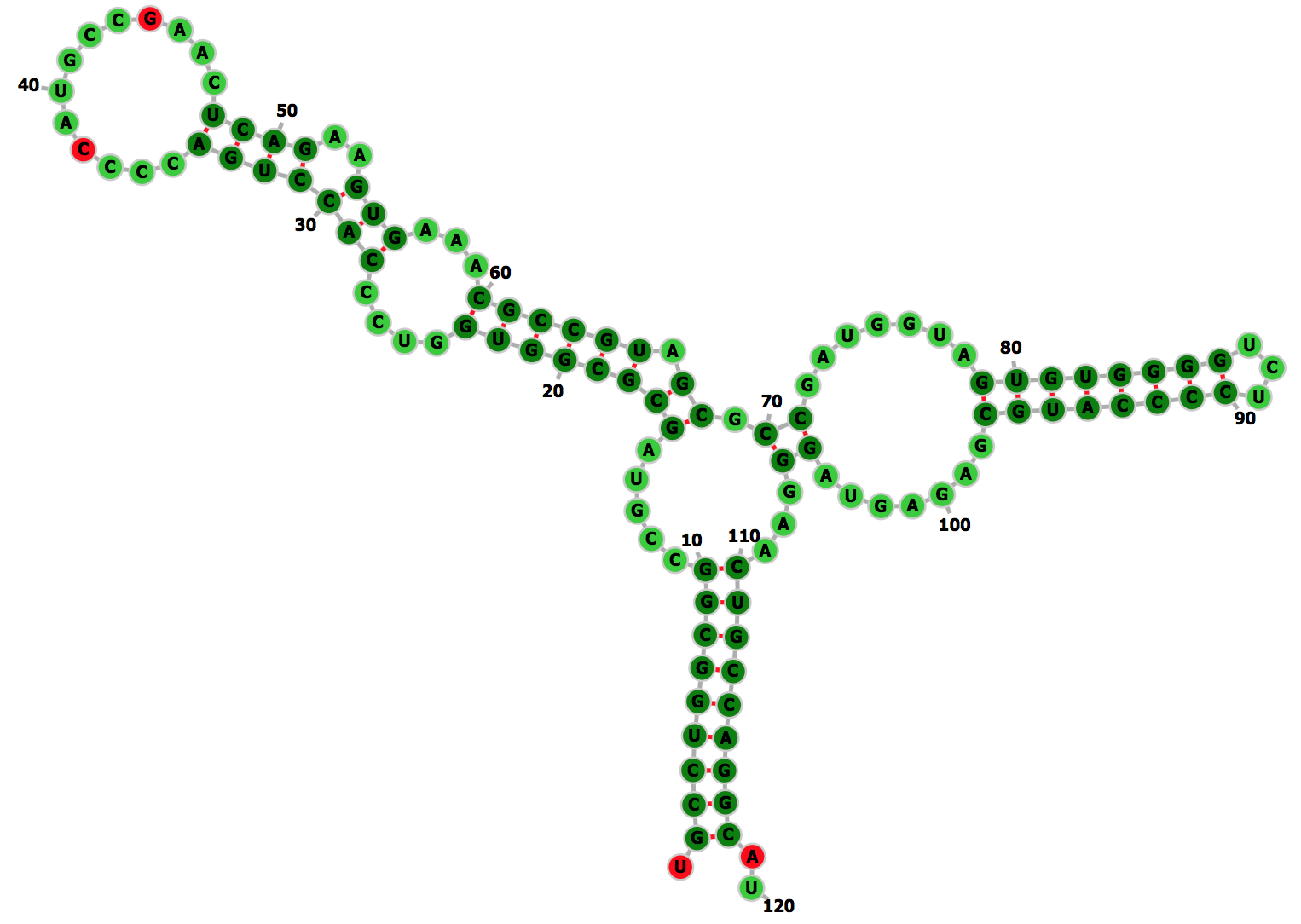}
        \caption{Trained model (0.95)}
    \end{subfigure}\vfill
     \begin{subfigure}[t]{0.33\textwidth}
        \includegraphics[width=1\linewidth,height=3.5cm, keepaspectratio]{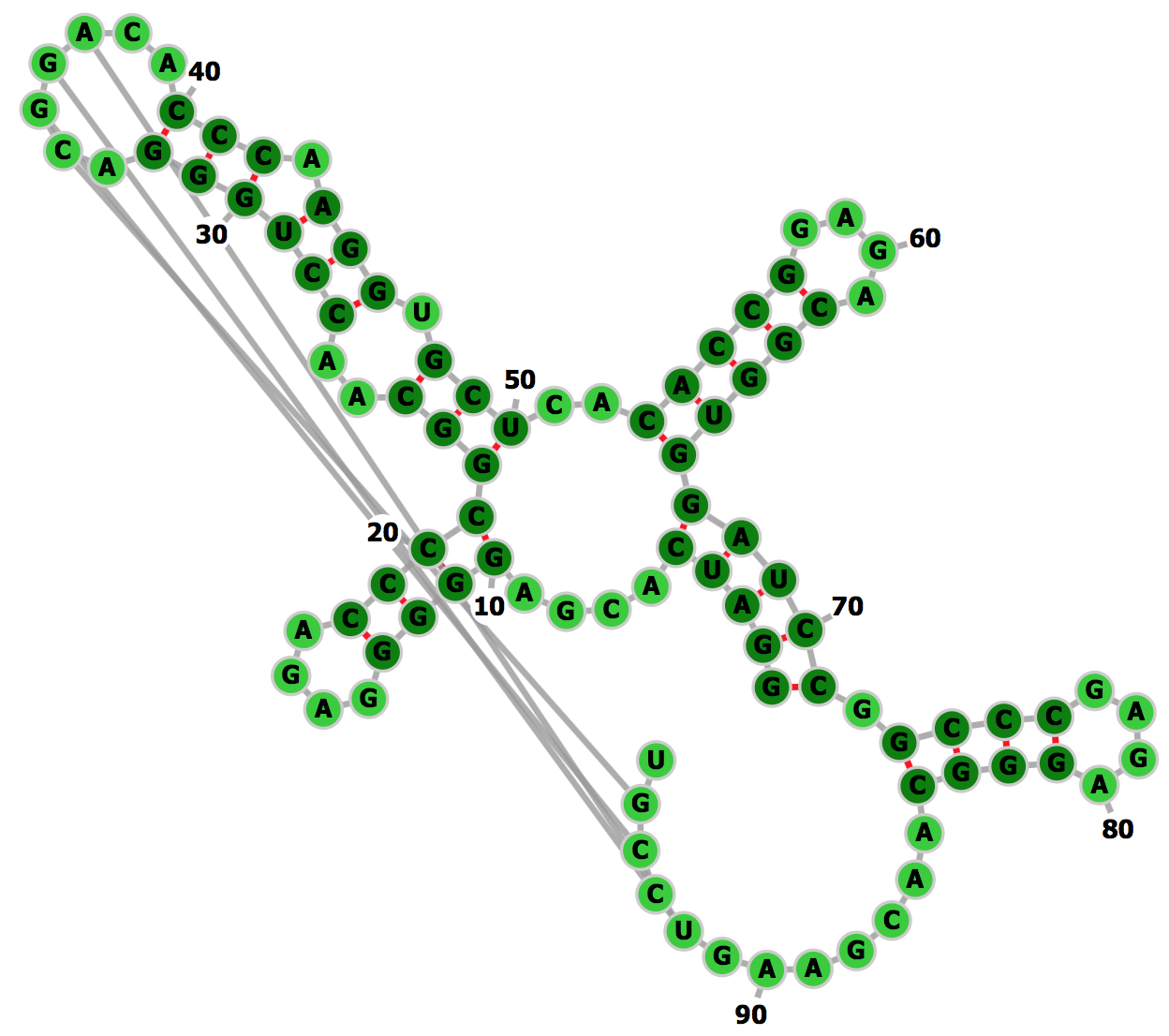}
        \caption{Reference}
    \end{subfigure}\hfill
    \begin{subfigure}[t]{0.33\textwidth}
        \includegraphics[width=1\linewidth,height=3.5cm, keepaspectratio]{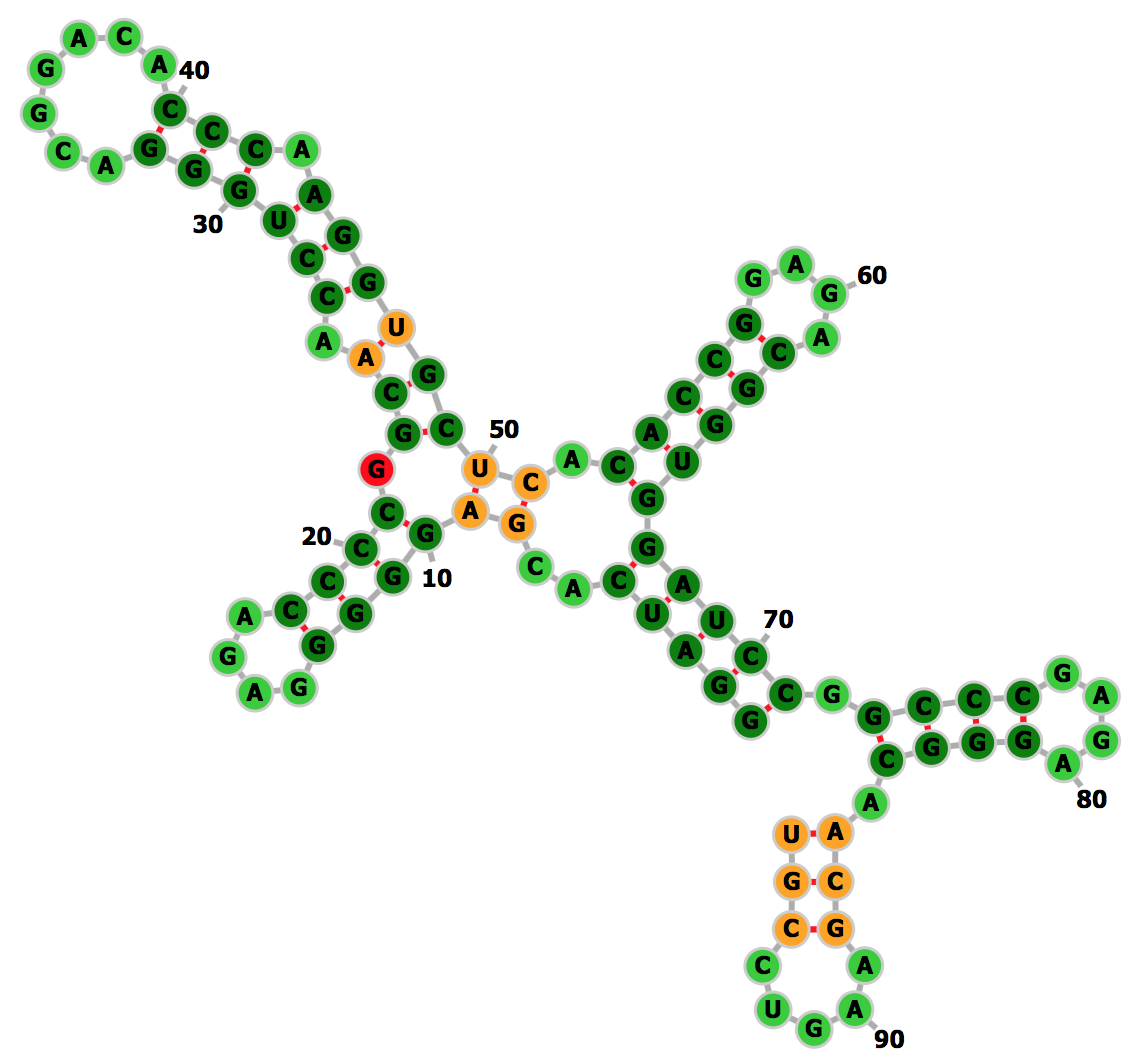}
        \caption{RNAfold (0.87)}
    \end{subfigure}\hfill
     \begin{subfigure}[t]{0.33\textwidth}
        \includegraphics[width=1\linewidth,height=3.5cm, keepaspectratio]{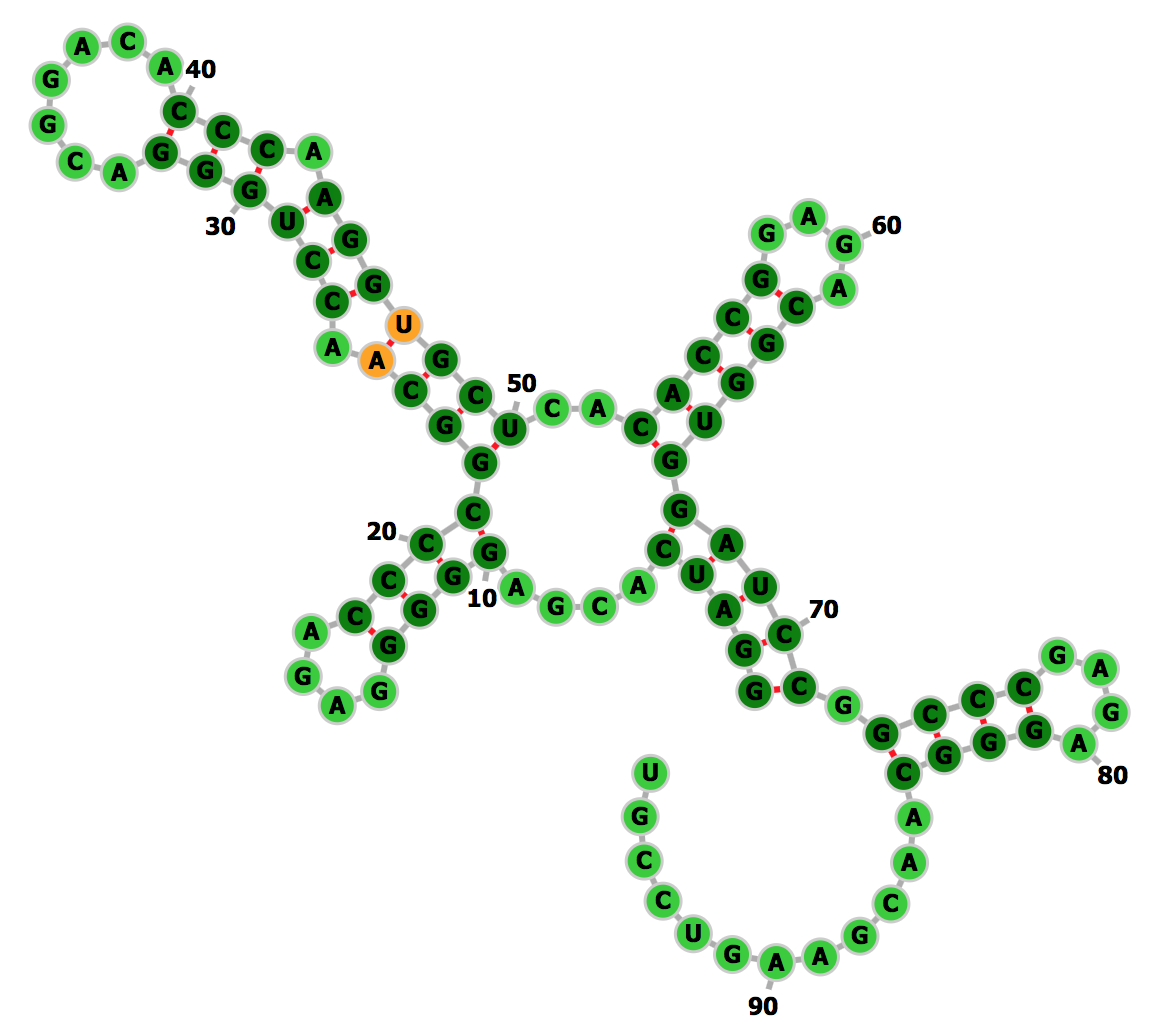}
        \caption{Trained model (0.98)}
    \end{subfigure}\vfill
     \begin{subfigure}[t]{0.33\textwidth}
        \includegraphics[width=1\linewidth,height=3.5cm, keepaspectratio]{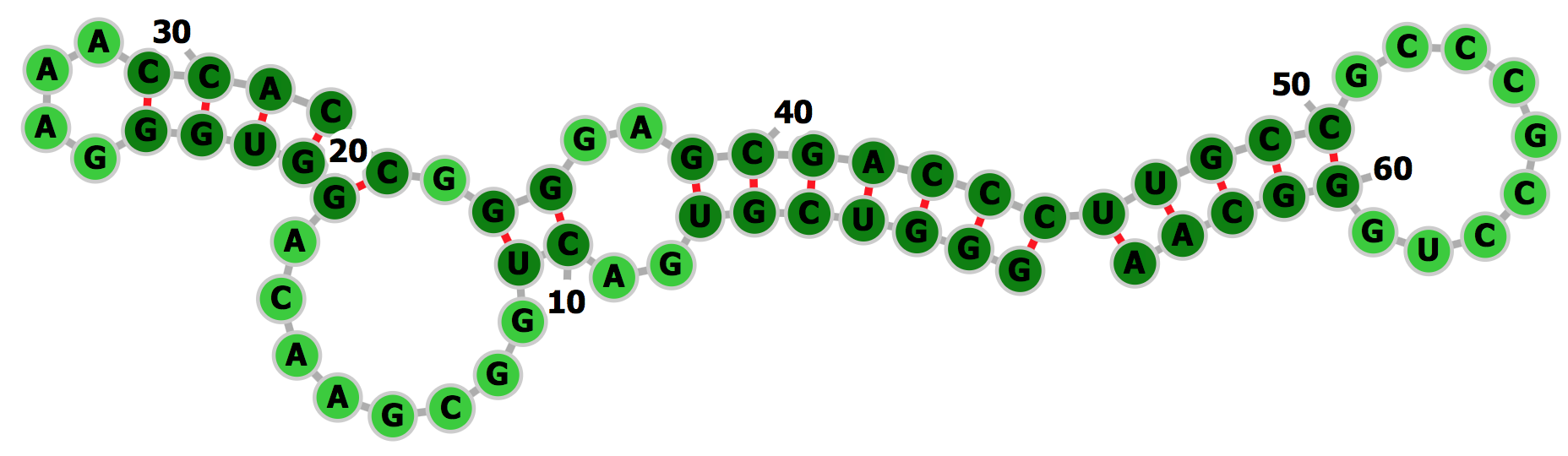}
        \caption{Reference}
    \end{subfigure}\hfill
    \begin{subfigure}[t]{0.33\textwidth}
        \includegraphics[width=1\linewidth,height=3.5cm, keepaspectratio]{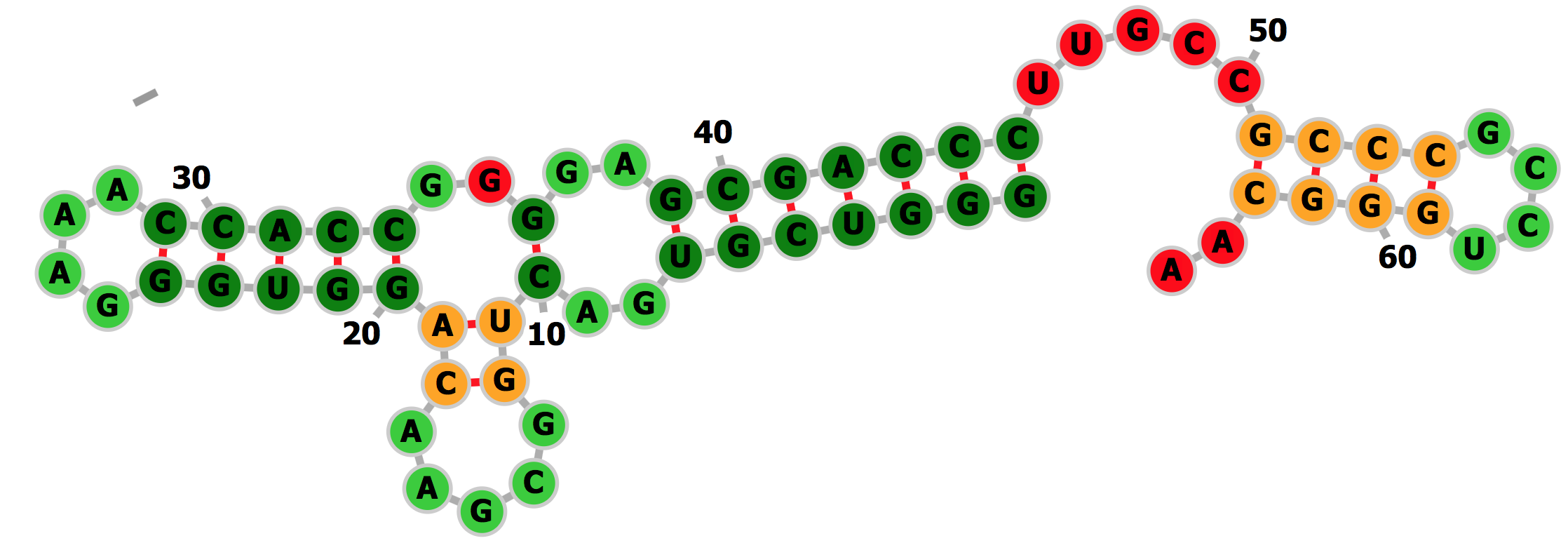}
        \caption{RNAfold (0.65)}
    \end{subfigure}\hfill
     \begin{subfigure}[t]{0.33\textwidth}
        \includegraphics[width=1\linewidth,height=3.5cm, keepaspectratio]{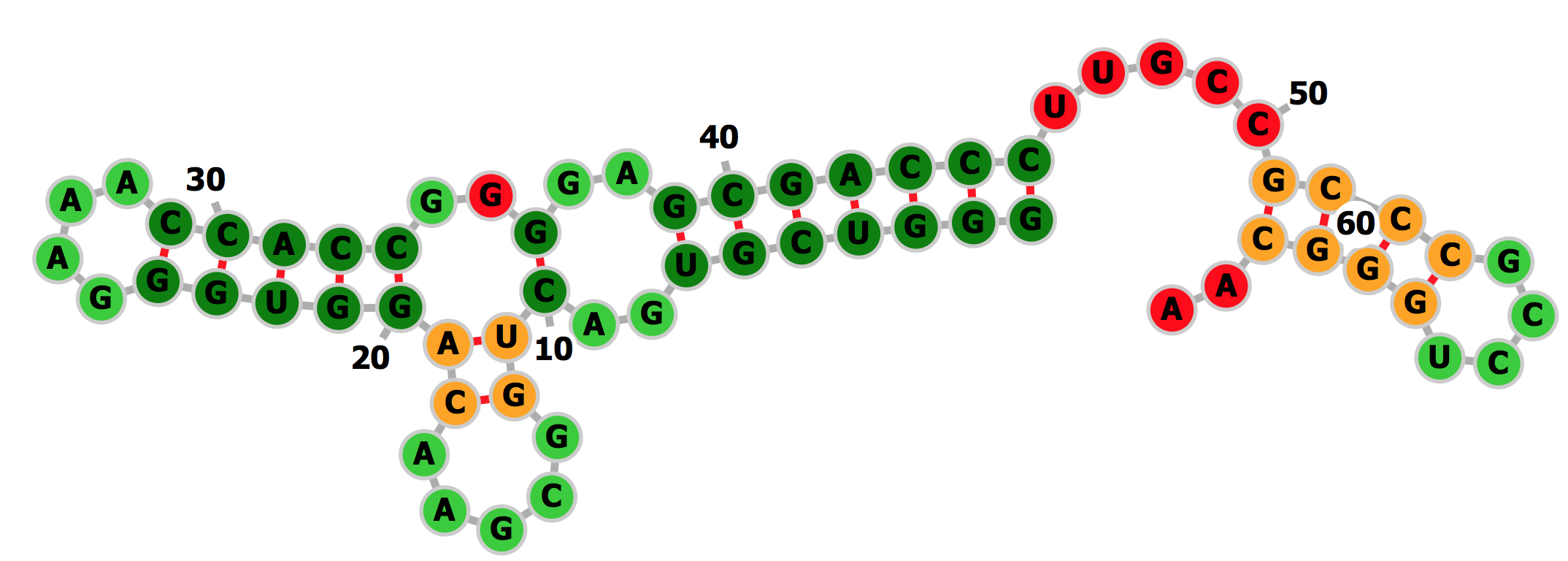}
        \caption{Trained model (0.65)}
    \end{subfigure}

\caption{Minimum free energy (MFE) structure predictions. For each system in the validation set,
        reference native structure is compared with predicted MFEs. Correctly predicted base pairs
        (true positives) and unpaired nucleotides (true negatives) are reported in dark green and
        lime green, respectively. Wrongly predicted base pairs (false positives) and unpaired nucleotides
        (false negatives) are reported in orange and red, respectively. 
        MCC between prediction and reference is reported in parenthesis. 
        All the relevant improvements in the prediction of these structures are reported in detail
        in section Results. 
        All secondary structure diagrams are drawn with \texttt{forna}~\cite{kerpedjiev2015forna}.
     }
    \label{mcc}
\end{figure*}

For the 6 test systems (splitting S3 of Table \ref{dataset}), the introduced procedure
leads to a boost of the population of the native structure by $\approx$ 19 times, on average 
(Fig.~\ref{populations}a, right side of the vertical line), when using the selected model $\{\alpha_S=0.001,\alpha_D=0.001,p=0\}$.
A side effect of targeting the population of native structures for model optimization and
selection is the increase in the similarity between the predicted minimum free energy (MFE)
and the experimental structures. This similarity can be quantified using the
Matthews Correlation Coefficient (MCC) \cite{matthews1975mcc}, that is routinely used to benchmark
RNA structure prediction \cite{miao2017rna}. Its average on the validation set is increased from $0.68$ to $0.89$ (Fig.~\ref{populations}d, right side).
Specific changes in the predicted secondary structures are reported in detail in Fig.~\ref{mcc}, 
where reference secondary structures are compared with MFE predictions made with unmodified RNAfold and with the selected model.
In particular, for \texttt{2GDI} (Fig.~\ref{mcc}a-c) our model recovers the correct structure of the 3-way junction loop (4-5:41-47:72-75);
for \texttt{2GIS} (Fig.~\ref{mcc}d-f) it recovers the correct structures of the hairpin loop (23-29) and the 
internal loop (17-21:31-38);
for \texttt{3DIG} (Fig.~\ref{mcc}g-i) the correct bulge loop (84-85:109-111) is recovered;
for \texttt{4YBB} (Fig.~\ref{mcc}j-l) the bulge loops (30-31:51-54) and (17-18:65-67), the internal loops (23-28:56-60)
and (71-79:97-105), the 3-way junction (10-16:68-70:106-110) and the hairpin loop (86-90);
for \texttt{4L81} (Fig.~\ref{mcc}m-o) the 4-way junction (5-10:21-22:51-53:66-67) is correctly predicted; for \texttt{4XW7} (Fig.~\ref{mcc}p-r) we have no change in MFE prediction with respect to unmodified RNAfold. 
Considering all of the tested splittings of the dataset,
the average MCC of minimum free energy structure predictions is increased from $0.72\pm 0.22$ to $0.90\pm 0.10$, implying both an increased average and a decreased variance
(details in Supporting Information).

As can be seen from Fig.~\ref{mcc}, some of the structures in the dataset contain pseudoknots. This kind
of pairing is forbidden in RNAfold structure predictions, thus we do not include it in the estimation of
MCC. Nontheless, data from both chemical probing and coevolution analysis in principle
contain information about pseudoknots, and it is possible to examine how reactivities and DCA scores
of pseudoknotted nucleotides are mapped into pairing penalties in our optimal model. 
We notice that the average value of pairwise DCA penalties applied to pseudoknot pairs $\langle\lambda_{ij}\rangle_{PK}=-0.087$ 
is comparable to the average of those applied to base pairs $\langle\lambda_{ij}\rangle_{BP}=-0.094$, so that
they have almost the same effect in favouring pairing (free-energy term $\lambda_{ij}s_{ij}<0$ for $s_{ij}=1$). 
The difference between the two values is negligible when compared with the average DCA penalty applied to unpaired 
nucleotides $\langle\lambda_{ij}\rangle_{UP}=0.447$. Reactivity-driven single-point penalties favour
unpaired states on average (free-energy term $-\lambda_{i}s_{i}>0$ for $s_{i}=1$), but the effect
on pseudoknotted nucleotides $\langle\lambda_i\rangle_{PK}=-0.142$ and on base-paired nucleotides
$\langle\lambda_i\rangle_{BP}=-0.125$ is approximatley half of that on unpaired nucleotides $\langle\lambda_i\rangle_{UP}=-0.284$.            
Even though in our optimal model the pairing of pseudoknotted nucleotides is boosted with almost the same intensity of base-paired nucleotides,
eventually small values are predicted for the corresponding pairing probabilites (see Supporting Information).
This is due to the fact that the thermodynamic model only allows structures with nested pairs.
 
It is also possible to test the scenarios where only DCA data or only chemical probing data are available.
In scenarios where only DCA information is used ($\alpha_S=\infty$), the best performance in CV is obtained using the model
with $\alpha_D=0.0001$ ($10\times$ increase in population, average MCC $=0.83$, Fig.~\ref{populations}b and d). 
This model is thus transferable to the validation set yielding a significant increase in both the 
population of the native structures and in MFE structure accuracy.
In the case of chemical probing-only information ($\alpha_D=\infty$), the best performance in CV is obtained using the model 
with hyperparameters $\alpha_S=0.01$ and $p=0$ ($3\times$ increase in population, average MCC $=0.71$, Fig.~\ref{populations}c and d).
Interestingly, whereas reactivity-only models perform systematically better in training than DCA-only models, their performance in CV is  
systematically lower, suggesting a lower transferability to unseen data, and thus a larger risk for reactivity-driven penalties to be overfitted.
This might be related to the high heterogeneity of the chemical probing data used here, that makes it difficult to fit transferable parameters.

Our procedure to compute pairing penalties from SHAPE reactivities can be compared with the one introduced by Deigan et. al. \cite{deigan2009accurate}.
Since the Deigan's method requires SHAPE data normalized with a different procedure,
we use normalized reactivities reported in Ref.~\cite{loughrey2014shape}. Remarkably, our procedure leads to
significantly better results both for molecules that are included in the training set (\textit{e.g.}, \texttt{4P8Z}, \texttt{1EHZ},
\texttt{5KPY},\texttt{1Y26},\texttt{4QLM} in Fig.~\ref{populations}c), 
and for most of the RNAs included in the validation set (\texttt{4YBB\_CB} and \texttt{4L81} in the right side of Fig.~\ref{populations}c).

\subsection{Interpretation of parameters}
\label{sec:interpretation}
\begin{figure*}
\centering
    \begin{subfigure}{0.33\textwidth}
        \includegraphics[width=1\linewidth]{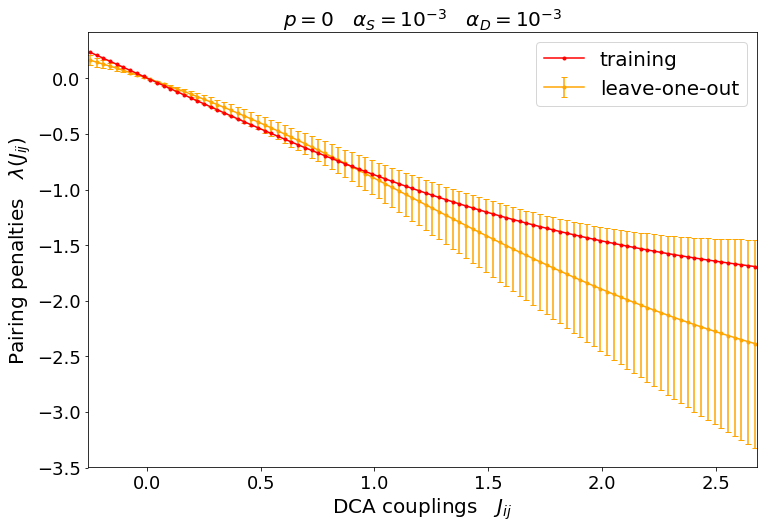} 
        \caption{}
    \end{subfigure}\hfill
    \begin{subfigure}{0.33\textwidth}
        \includegraphics[width=1\linewidth]{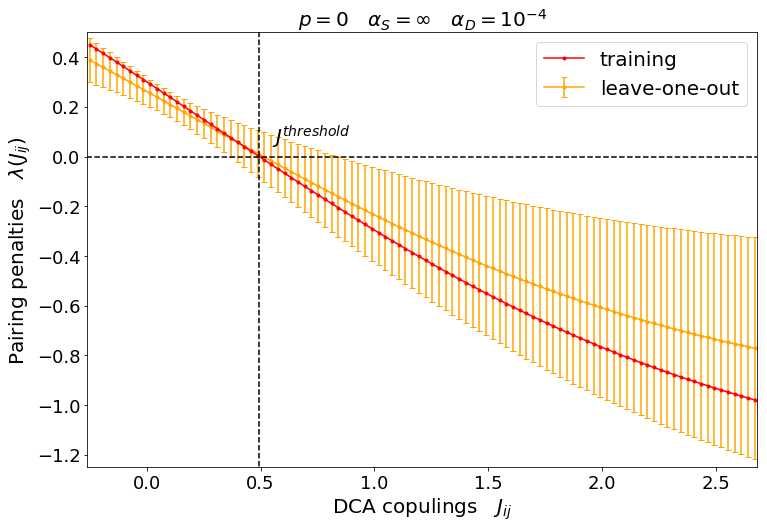} 
        \caption{}
    \end{subfigure}\hfill
     \begin{subfigure}{0.33\textwidth}
      \includegraphics[width=1\linewidth]{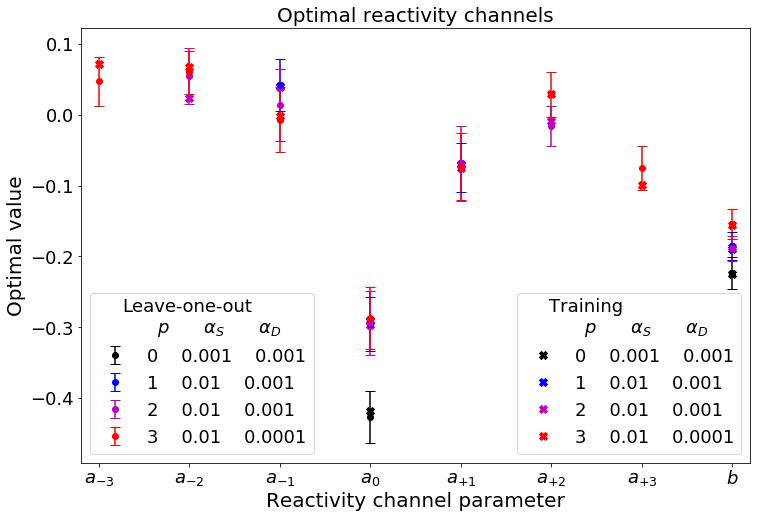}
        \caption{}
    \end{subfigure}
    \caption{Properties of the optimized neural network. 
    For the DCA channel, the optimized function mapping DCA couplings $J_{ij}$ into pairing penalties $\lambda_{ij}$,
    for both (a) the selected model and (b) the best performing model with restriction to only DCA input. When trained 
    on the whole training set (red) the activation function is consistent with the average on the leave-one-out training 
    subsets (orange). Error bars are computed as standard deviations and are significantly lower in the region of DCA couplings
    around zero, as couplings lying in that region are more frequent. The trained function maps high (respectively, low) DCA coupling values to 
    penalties favoring (respectively, disfavoring) the corresponding pairings, thus affecting the population of the structures including the specific pair. 
    When restricting to (b) models including only DCA input, the threshold value of the coupling $J^{threshold}$ between
    disfavored and favored pairing corresponds to the zero of the activation function, as indicated by the dashed line. 
    For the chemical mapping channel, (c) optimal values of model parameters are shown for the selected model (black)
    with hyperparameters $\{\alpha_S=0.001,\alpha_D=0.001,p=0\}$, and
    for the sub-optimal models with $p>0$. All the training results (cross) lie within the leave-one-out errors (dots with error bars), 
    indicating robustness of the minimization procedure against cross-validation. 
    Coefficients $\{a_{-k},\dots,a_{+k}\},\quad k>0$ weighting reactivities up to the $k$-th nearest-neighbors
    of a nucleotide, report the minor contributions of the local reactivity pattern in addition to 
    the nucleotide's own reactivity.
}
    \label{dca_sigmoid}
\end{figure*}
In principle, different randomizations of the training set yield different hyperparameters and parameters
for the functions implemented in the selected model.
Here we continue focusing on splitting S3.
The selected model is defined by hyperparameters $\{\alpha_S=0.001,\alpha_D=0.001,p=0\}$. 
Results for different splittings are similar and are reported in Supporting Information.

\emph{DCA channel.}
DCA couplings are mapped into pairing penalties through a double-layered neural network,
resulting in a non-linear function reported in Fig.~\ref{dca_sigmoid}a.  Pairing penalties are found to
decrease with increasing DCA coupling value, consistent with the interpretation that
large couplings should correspond to co-evolutionarily related and thus likely paired nucleobases \cite{de2015direct}.
A more detailed interpretation of these pairing penalties is possible if we restrict to models taking only DCA couplings as input ($\alpha_S=\infty$).
The corresponding non-linear function is reported in Fig.~\ref{dca_sigmoid}b. The overall shape
is consistent with that obtained fitting all the data (Fig.~\ref{dca_sigmoid}a),
but the zero of this function can be straightforwardly interpreted as the threshold for penalizing or favoring base pairing.
The resulting value is $J^{\textrm{threshold}}=0.49$ consistent with the typical thresholds obtained in \cite{cuturello2018assessing}
with a different optimization criterion, based on the accuracy of contact predictions, and fitted on a larger 
dataset, thus confirming the transferability of the non-linear function reported here.

\emph{Reactivity channel.}
Chemical probing reactivities are mapped into penalties affecting the population of individual nucleotide pairing states
through a single convolutional layer with a linear activation function.
When evaluated on the training set, the best performance is obtained with models including up to the maximum tested number 
of nearest neighbors ($p=3$).
In these models, for each nucleotide, the network input vector includes reactivities from the third nearest-neighbor upstream
to its third nearest-neighbor downstream along the sequence.
The activation coefficients $\{a_k,\quad k=-3,\dots,+3\}$ weight the contribution of each nucleotide in the neighbor window.
Despite the performance improvement on the training set, transferability to data not seen during the training phase is 
best preserved in the model that retains 
only the contribution from the $a_0$ term, confirming that the reactivity 
of a nucleotide is maximally affected by its pairing state.
We notice that SHAPE reactivity has been correlated with sugar flexibility \cite{weeks2011exploring,mlynsky2018molecular,frezza2019interplay},
and is only indirectly related to the pairing state of a nucleotide. Nevertheless, reactivity information can be used to systematically improve
predictions at the base-pairing level.
In particular, $a_0 < 0$ (see Fig.~\ref{dca_sigmoid}c, black)  
so that the pairing of a highly reactive nucleotide is unfavoured and vice-versa for nucleotides with low reactivity.
On the other hand, the best (suboptimal) neighbor-including models (\textit{i.e.}, with $p > 0$) still yield comparable results
with respect to the selected one and significant improvements as well with respect to thermodynamic model alone.
Figure~\ref{dca_sigmoid}c reports the sets of optimal parameters with $p > 0$.
We notice that at each increment of $p$, when two new parameters $a_p$ and $a_{-p}$ are introduced,
all the shared subsets $\{a_{p-1},\dots ,a_{-p+1}\}$ overlap significantly, and a number of features are shared as well.
First, for all the optimal choices of $p > 0$, the sum of the weights $\sum_{i=-p}^{p} a_i$ is negative, so that the pairing of a nucleotide in
a highly reactive region is unfavored, and vice-versa for regions of low reactivity.
The largest contribution still arises from the $a_0$ term, but it is slightly lower in absolute value, to compensate for neighbor corrections.
For each pair (downstream and upstream) of $k$-th nearest-neighbours, the combination of the $a_0$ and $a_{+k}$ ($a_{-k}$) contributions can be interpreted as a
forward (backward) finite-difference operator estimating the $k$-th order derivative of the reactivity with respect to the
position in the sequence. These contributions map local downward trends of the reactivity profile
into pairing penalties, thus providing a sort of normalization for the reactivity of the central nucleotide with respect to that of its neighbors.
As the order of the derivatives increase from the first, weights become lower such that the corresponding corrections progressively decrease
in importance. It is interesting to notice that the finer these corrections are, the more the corresponding parameters tend to be overfitted
to the training set.

\section{DISCUSSION}

In this work we build a network that can be used to predict RNA structure
taking as an input RNA sequence, chemical probing reactivities, and DCA scores.
Whereas reactivities and DCA scores are processed through standard linear or non-linear units,
RNA sequence enters through a thermodynamic model.
A crucial ingredient that we introduce here are the derivatives of the result of the thermodynamic model with respect to the
pairing penalties, that allow the network to be trained using gradient-based machine learning techniques.

We built up a total of 196 models to map simultaneously reactivities and DCA scores into free-energy terms coupling, respectively, 
the pairing state of individual nucleotides and that of specific pairs of nucleotides. Each model is defined by
tunable hyperparameters controlling the width of the windows used to process reactivities and the
strength of the regularization terms applied to chemical probing and DCA data.
The dataset is \textit{a priori} split randomly into a training set and a validation set (12 and 6 systems respectively).
Training, model selection and validation are repeated for different random splittings of the dataset, in order to 
decrease the chance of introducing a bias towards specific structures or features, and ensuring the robustness 
of the procedure. 
The whole procedure, from training to model selection, is automatic so that new parameters
could be straightforwardly obtained using new chemical probing and DCA data and new crystallographic structures,
allowing for a continuous refinement of the proposed structure prediction protocol\textcolor{blue}.
Training one model required 20 minimizations that were performed in parallel on nodes containing 2 E5-2683 CPU each, using 20 cores. 
Each minimization took approximately 30 minutes, though the exact time depends on the value of $p$ and on the system size.
4x7x7=196 minimizations were done to scan the hyperparameter space. 12 separate models needed to be trained for the leave-one-out. 
Notably, the dependence between the minimizations (see Materials and Methods) can be taken into account (see scripts in Supporting Information) allowing 
them to be largely run in parallel. In practice, if 288 nodes are simultaneously available, the full minimization for 12 systems 
can be run in approximately 8 hours.
In the dataset we used, some reactivities are taken from available experimental data.
Other reactivities are measured here for the first time so as to increase the number of systems for which
both co-evolutionary data and reactivities are available.
DCA scores are based on ClustalW alignments \cite{larkin2007clustal} so that they are not manually curated with prior structural 
information. We notice however that classification of sequences in RFAM is performed including structural information,
when available.
In addition, co-evolutionary information might be difficult to extract for poorly conserved long non-coding RNAs.
All the reactivity profiles and DCA score matrices are reported in Supporting Information Section S2.
All the results obtained with different randomization of the validation set are reported in Supporting Information
so that different sets of parameters can be easily tested.

The model selected via CV is defined by hyperparameters $\{p=0,\alpha_S=0.001,\alpha_D=0.001\}$. 
The best performance/transferability trade-off is thus obtained when not incorporating reactivities from neighboring
nucleotides in the pairing state of a nucleotide.
This model is systematically capable of predicting a higher population for the native structure.
The model that is selected using only chemical probing data yields better results in population than what obtained with Deigan's method
\cite{deigan2009accurate}, which is accounted for best state-of-the-art method \cite{lorenz2016shape} among those
based on SHAPE reactivities only.
Results obtained with our selected model confirm that the reactivity of a nucleotide is a good indicator of its own pairing state \cite{lorenz2016shape}.
We also observe that the reactivity of neighbors correlate too with the pairing state of a nucleotide (see Supporting Information).
However, the pairing state of neighboring nucleotides is implicitly taken into account in the RNAfold model, that includes energetic contributions
for consecutive base pairs, implying that the explicit inclusion might not be required. More precisely, the need for a larger number
of parameters to be trained when increasing the $p$ hyperparameter might not be compensated by a sufficient improvement in the prediction performance.
Interestingly, in a previous version of this work based on a smaller dataset and on different thermodynamic parameters \cite{xia1998thermodynamic}
the most transferable model identified had $p=2$
(see \url{https://arxiv.org/abs/2004.00351v1}).
In perspective, the model can be extended to include additional features of the chemical probing experiments that may be related to non-canonical
interactions and three-dimensional structure.  

Although our selected model is trained to maximize the population of the individual reference structure
as obtained by crystallization experiments, it can still report
alternative structures. Whereas we did not investigate this issue here, alternative low-population states might be highly
relevant for function. Compatibly with that,
the absolute population of the native structure remains significantly low
(from $\approx 10^{-8}$ to $\approx 10^{-7}$), but is still one of the highest in the ensemble.
In particular, the individual structure with highest population (minimum free-energy structure)
with our method is closer to the reference crystallographic structure than the one predicted by thermodynamic parameters alone
on systems not seen during training.

Importantly, all the data and the used scripts are available and can be used to fit the model over larger datasets.
In order to avoid overfitting, we suggest to repeat the leave-one-out procedure to select the most 
transferable model, whenever new independent data is added to the dataset. Scripts for training
and model selection are reported in Supporting Information.  
In principle the model can be straighforwardly extended to include any chemical probing data that
putatively correlates with base pairing state \cite{weeks2010advances}
or other types of experimental information that correlate with base-pairing probabilities
\cite{ziv2018comrades}.
Training on a larger set of reference structures and using more types of experimental data
will make the model more robust and open the way to the reliable structure determination of non-coding RNAs.

\end{document}